\documentclass[aps,pre,preprint,superscriptaddress]{revtex4}
\usepackage{amsmath, amssymb}
\usepackage{graphicx}
\usepackage{latexsym}
\usepackage{esint}
\usepackage{cancel}

\begin{document}

\title[The effect of feedback in shaping receptive fields]{Effects of spike-triggered negative feedback on receptive-field properties}

\author{Eugenio Urdapilleta}
\email[]{eugenio.urdapilleta@cab.cnea.gov.ar}
\affiliation{Divisi\'on de F\'isica Estad\'istica e
Interdisciplinaria, Centro At\'omico Bariloche, Avenida E.
Bustillo Km 9.500, S. C. de Bariloche (8400), R\'io Negro,
Argentina\\\it{Present address: Cognitive Neuroscience, SISSA, via
Bonomea 265, 34136 Trieste, Italy}}

\author{In\'es Samengo}
\affiliation{Divisi\'on de F\'isica Estad\'istica e
Interdisciplinaria \& Instituto Balseiro, Centro At\'omico
Bariloche, Avenida E. Bustillo Km 9.500, S. C. de Bariloche
(8400), R\'io Negro, Argentina}

\date{Received: date / Accepted: date}

\begin{abstract}
Sensory neurons are often described in terms of a receptive field,
that is, a linear kernel through which stimuli are filtered before
they are further processed. If information transmission is assumed
to proceed in a feedforward cascade, the receptive field may be
interpreted as the external stimulus' profile maximizing neuronal
output. The nervous system, however, contains many feedback loops,
and sensory neurons filter more currents than the ones representing
the transduced external stimulus. Some of the additional currents
are generated by the output activity of the neuron itself, and
therefore constitute feedback signals. By means of a time-frequency
analysis of the input/output transformation, here we show how
feedback modifies the receptive field. The model is applicable to
various types of feedback processes, from spike-triggered intrinsic
conductances to inhibitory synaptic inputs from nearby neurons. We
distinguish between the intrinsic receptive field (filtering all
input currents) and the effective receptive field (filtering only
external stimuli). Whereas the intrinsic receptive field summarizes
the biophysical properties of the neuron associated to subthreshold
integration and spike generation, only the effective receptive field
can be interpreted as the external stimulus' profile maximizing
neuronal output. We demonstrate that spike-triggered feedback shifts
low-pass filtering towards band-pass processing, transforming
integrator neurons into resonators. For strong feedback, a sharp
resonance in the spectral neuronal selectivity may appear. Our
results provide a unified framework to interpret a collection of
previous experimental studies where specific feedback mechanisms
were shown to modify the filtering properties of neurons.

\keywords{Receptive field \and Adaptation \and Feedback \and
Resonance}
\end{abstract}

\maketitle

\section{Introduction}\label{intro}
\begin{sloppypar}
\indent Sensory areas are exposed to large variations in the
physical magnitudes they encode \cite{DongAtick1995,
SchwartzSimoncelli2001, Geisler2008, Theunissen_etal2000}. The
dynamic range of input signals can often span several orders of
magnitude during the course of a single behaviorally relevant time
interval. For example, in the visual system, mean luminosity
changes drastically when the gaze is displaced from a spot that is
directly illuminated by the sun, to a shadowy corner. Neural
systems have therefore developed adaptive mechanisms, modifying
the neural code according to the sensory context
\citep{Ulanovsky_etal2003, Mante_etal2005}.

\indent Several types of adaptive mechanisms exist, as for
example, synaptic plasticity \cite{AtwoodKarunanithi2002,
Xu-FriedmanRegehr2004, Sjostrom_etal2008, Feldman2009}, feedback
through recurrent connectivity \cite{Douglas_etal1995,
Shu_etal2003, Eytan_etal2003, BuonomanoMaass2009}, feedback
through adaptation currents \cite{Wang1998, SanchezVives_etal2000,
PrescottSejnowski2008, PeronGabbiani2009}, and intrinsic
non-linear effects \cite{Borst_etal2005}. Different mechanisms
operate on different timescales; whereas non-linear mechanisms
emerge in a matter of milliseconds, synaptic plasticity usually
develops in one or a few seconds. Changes in the intermediate
range (hundreds of milliseconds) are mostly due to adaptation
currents and recurrent connectivity. At these time scales,
adaptivity arises from the dynamics associated to certain
processes at cellular and network levels, without the profound
reorganization entailed by learning and plasticity.

\indent Adaptive phenomena mediated by intrinsic currents and
feedback network connectivity are based on history-dependent
spike-evoked activity. They exert their feedback influence mainly
by reducing neuronal gain \cite{Felsen_etal2002, BendaHerz2003,
AyazChance2009, Benda_etal2010}, and thereby, by modifying the
input/output relation of the cell, that is, the relation between
spiking rate and mean stimulus strength. However, adaptation
phenomena go far beyond a mere reduction in firing rates, often
involving a dramatic reshaping of the selectivity to
time-dependent inputs, and of the statistics of neuronal output.
For example, in the olfactory bulb, feedback has been shown to
amplify input fluctuations of a specific frequency, and thereby to
induce strong oscillations in the output activity
\cite{Freeman1972a, Freeman1972b, Freeman1972b}, giving rise to
complex (often chaotic) dynamic behavior \cite{Freeman1987,
DavidFriston2003}. Similar conclusions have been reached in
theoretical explorations of interacting neural populations
\cite{WilsonCowan1972, CoombesLaing2009, Buice_etal2010,
Bressloff2012, AmitBrunel1997, LedouxBrunel2011}. Therefore,
although adaptation is usually claimed to have evolved in order to
increase the dynamic range of sensory encoding, its effect on the
temporal properties of the neural code should not be overlooked:
Feedback alters the basic properties of neuronal selectivity (as
explored below), and also the temporal evolution of the output,
the amount of temporal correlations, the precision and the
reliability of neural responses \cite{Ladenbauer_etal2014,
Urdapilleta2011, Butts_etal2011}.

\indent In the context of spike-evoked feedback, adaptation
processes have been shown to modify the selectivity to transient
stimulus temporal profiles, enhancing the representation of
high-frequency components \cite{BendaHerz2003, Gigante_etal2007,
BendaHennig2008, Benda_etal2010}. Such changes become evident when
computing the receptive field, or the relevant stimulus
directions, by means of reverse correlation techniques
\cite{DayanAbbott2001, Chichilnisky2001, SamengoGollisch2013}.
Several previous studies have demonstrated that adaptation in mean
firing rates may or may not be accompanied by changes in receptive
fields \cite{Enroth-CugellShapley1973, Victor1987, Butts_etal2011,
BaccusMeister2002, Sharpee_etal2011, GarvertGollisch2013}. To our
knowledge, there is yet no theoretical framework that allows us to
understand why and when receptive fields are expected to be
modulated by feedback. Such theoretical framework should be
general enough to be applicable to cases where adaptation is
mediated by feedback at the network level, or at the level of
voltage-dependent ionic conductances and refractoriness, as for
example, in the Hodgkin-Huxley model \cite{Samengo_etal2013} or in
a LGN model neuron \cite{GaudryReinagel2007}. In this work we
describe the adaptive changes observed in the stationary and
transient encoding properties of linear Poisson models driven by a
combination of external stimuli and spike-triggered negative
feedback. We theoretically analyze how adaptation modifies the
shape of the receptive field, providing a complete spectral and
temporal description in the limit where the feedback signal is
fully determined by the spiking probability of the cell under
study (perfect feedback). Spike-evoked negative feedback is shown
to induce divisive gain control, to reshape the receptive field,
and to enhance resonant properties. In order to extend these
results to the case where feedback is a noisy function of the
spiking probability, we incorporate stochastic elements into the
theory. Finally, we also discuss an extension to non-linear
Poisson models.
\end{sloppypar}

\section{Results}
\subsection{Theoretical description of perfect feedback}
\indent The main goal of this work is to study how the receptive
field of a neuron varies, when feedback processes are incorporated.
In addition to the temporal dimension, receptive fields may be
defined as a function of a variety of additional dimensions
(spatial, frequency, chromatic, chemical) depending on the modality
of the sensory system under study (vision, audition, taste, etc.).
Since adaptation processes unfold along the temporal domain, for the
moment, we restrict the analysis to the temporal profile of
receptive fields, and defer to the last section of result the
extension to higher-dimensional problems.

\indent Strictly speaking, the concept of receptive field is well
defined for linear-nonlinear Poisson models. In order to develop
the theoretical framework, we initially restrict to purely linear
Poisson models, and later on discuss the extension to the
nonlinear case. In the linear case, the probability of generating
a spike in the interval $[t, t + {\rm d}t]$ under the influence of
stimulus $I(t)$ is $r(t) \ {\rm d}t$, with

\begin{equation} \label{equation_1}
  r(t) = h_0 + \int_{-\infty}^{\infty} h(\tau)~I(t-\tau)~{\rm d}\tau.
\end{equation}

\indent Here, $h_0$ is the spontaneous firing probability, and
$h(\tau)$ is the receptive field of the cell. Since only past
stimuli can influence the present firing probability, causality
imposes that $h(\tau)$ = 0, for all $\tau < 0$. The stimulus
$I(t)$ can be interpreted as either the external signal controlled
by the experimentalist (light, sound, touch, etc.), or as the
input ionic current entering the cell. By stimulating the cell
with stochastic input signals, the shape of $h(\tau)$ may be
easily obtained through reverse correlation techniques
\cite{Chichilnisky2001, SamengoGollisch2013}. The shape, units and
dimensionality of the filter $h(\tau)$ naturally depend on whether
the input signal used in the reverse correlation analysis is the
macroscopic external stimulus, or the microscopic ionic current.
We specifically distinguish between the component of $I(t)$
representing the transformation of the external signal
accomplished by upstream neurons, and the component describing all
negative feedback signals that are triggered by previous activity
of the cell under study

\begin{equation} \label{equation_2}
  I(t) = s_0 + s_1(t) - g \ x(t).
\end{equation}

\indent In Eq.~(\ref{equation_2}), $s_0 + s_1(t)$ is the extrinsic
stimulus component stemming from the transduction of sensory
signals, and $x(t)$ is the feedback signal whose value depends on
the spiking history of our neuron. The coupling constant $g$ has
dimensions of transduced stimulus, and represents the strength of
the feedback connection. It may be either negative or positive,
depending on whether the cell under study fires in response to
positive or negative stimulus deflections. The sign must be chosen
in such a way as to produce a signal that opposes the natural
excitability of the cell, in order to avoid positive-feedback
instabilities (see below). We separate the average external
stimulus $s_0$, so that the time-dependent component $s_1(t)$ can
be assumed to have zero mean. Defining the baseline firing rate

\begin{equation} \label{equation_3}
  r_0 = h_0 + H \ s_0,
\end{equation}

\noindent and replacing Eq.~(\ref{equation_2}) in
Eq.~(\ref{equation_1}), we obtain

\begin{equation}\label{equation_4}
  r(t) = r_0 + \int_{-\infty}^{\infty} h(\tau)~\left[s_1(t - \tau) -
  g \ x(t - \tau) \right]~{\rm d}\tau,
\end{equation}

\noindent where

\begin{equation}\label{equation_5}
  H = \int_{0}^{\infty} h(\tau) \ {\rm d}\tau = \sqrt{2 \pi} \
  \hat{h}(\omega = 0).
\end{equation}

\indent The parameter $H$, hence, is proportional to the
ze\-ro-fre\-quen\-cy component of the Fourier transform of the
receptive field, $\hat{h}(\omega = 0)$ (the convention used for
the Fourier transform is specified in Appendix \ref{Apendice1}).

\indent We assume that the feedback signal $x(t)$ is boosted by
discrete pulses, and has a natural decay time $\tau_{\rm d}$. For
example, if feedback is implemented through the inhibitory action
produced by nearby neurons, the pulses represent spikes produced
by other cells in the network providing inhibitory feedback to the
neuron under study. Hence,

\begin{equation}\label{equation_6}
   \frac{{\rm d}x}{{\rm d}t} = - \frac{x}{\tau_{\rm d}} + \sum_{{\rm
   i}=1,~\forall {\rm k}}^{N} \alpha_{\rm i}~\delta(t-t_{\rm i}^{\rm k}),
\end{equation}

\noindent where $\alpha_{\rm i}$ weighs the increase of feedback
activity due to a spike in the $i$-th presynaptic neuron, occurred
at time $t_{\rm i}^{\rm k}$ (k collectively represents all spike
times). If Eq.~(\ref{equation_6}) is meant to represent a feedback
signal, the average activity of the neurons contributing to the
sum must be proportional to the output of the neuron under study.
Only with such proportionality can we ensure that $x(t)$ is linked
to the past activity of the neuron. For the sake of simplicity, we
model the link as a simple proportionality. Moreover, in our first
attempt to model feedback, we assume that the sum in
Eq.~(\ref{equation_6}) is not proportional to the actual output of
the cells, but rather, to the {\em probability} $r(t)$ to generate
a given output. That is,

\begin{equation}\label{equation_7}
  \sum_{{\rm i}=1,~\forall {\rm k}}^{N} \alpha_{\rm i}~\delta
  (t-t_{\rm i}^{\rm k}) \approx r(t).
\end{equation}

\indent This approximation is here called a \textit{perfect}
feedback signal, since $x(t)$ is a deterministic function of
$r(t)$; more specifically, it is a leakily integrated copy of
$r(t)$. Later on we discuss the case where the feedback signal is
a stochastic (as opposed to deterministic) function of $r(t)$, and
more accurately describes the actual output of the cell under
study. The approximation of perfect feedback is valid if there is
a large number of neurons contributing to the sum in
Eq.~(\ref{equation_6}), and if all of them have similar
statistical and dynamical properties, so that they are all
governed by the same firing probability. Alternatively, one may
assume that the external input drives a pool of independent
neurons, which collectively provides a normalizing signal
associated to the common processing of the incoming stimuli
\cite{Heeger1992, Carandini_etal1997, CarandiniHeeger2012}.
Mathematically, these conditions mean to assume that all the
$\alpha_i$ are equal, to take $N \to \infty$, and in order to
maintain the total input bounded, additionally scale $\alpha_{\rm
i} = 1 / N$. With these approximations, replacing
Eq.~(\ref{equation_7}) in Eq.~(\ref{equation_6}), we obtain

\begin{equation}\label{equation_8}
 \frac{{\rm d}x}{{\rm d}t} = - \frac{x}{\tau_{\rm d}} + r(t).
\end{equation}

\indent Equations~(\ref{equation_4}) and (\ref{equation_8})
constitute a closed set: If the external stimulus $s_0 + s_1(t)$
and the filter $h(\tau)$ are known, both the feedback term and the
firing probability can be calculated.

\indent The aim of this study is to deduce how feedback affects
the filtering characteristics of the neuron. To that end, we now
assume that both the external stimulus and the firing probability
are known (the latter can be recorded from repeated presentations
of the same stimulus), and evaluate whether the input/output
relation can still be considered a filtering process, in spite of
feedback. The linear nature of Eqs.~(\ref{equation_4}) and
(\ref{equation_8}) calls for a treatment in Fourier space. Based
on the properties of the Fourier transform (see Appendix
\ref{Apendice1}) and rearranging terms, we obtain

\begin{figure*}[t!]
\begin{center}
\includegraphics[scale = 1.00, angle = 0]{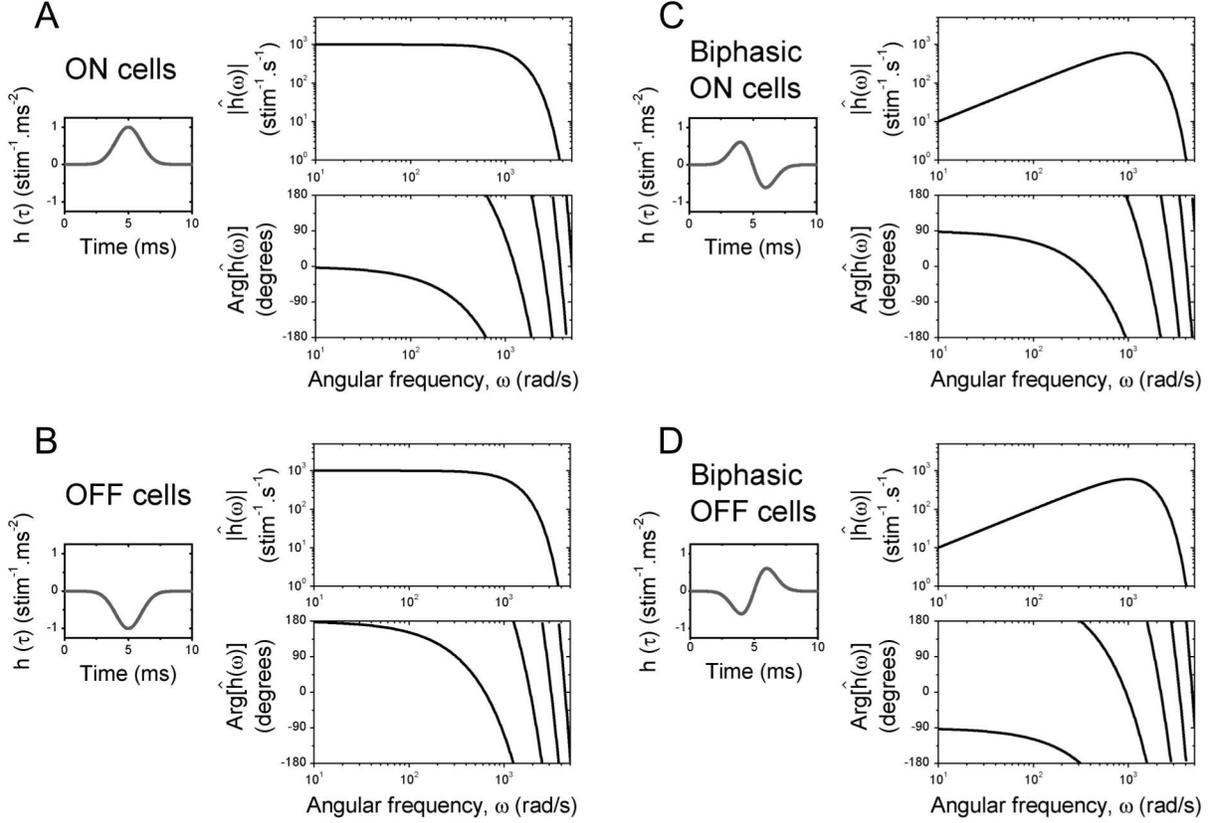}
\caption[Spectral characteristics of simplified ON, OFF, and
biphasic filters.]{\label{fig1} \linespread{1} Spectral
characteristics of simplified ON, OFF, and biphasic filters. {\sl
A}: Example of an ON filter displayed as a function of time
(left). The modulus and phase of its Fourier transform are
displayed as a function of frequency (right). Note the logarithmic
and semi-logarithmic scales. {\sl B}, {\sl C}, {\sl D}: Analogous
representations of OFF, biphasic-ON and biphasic-OFF filters,
respectively. In these examples, ON and OFF filters have exactly
the same $|\hat{h}(\omega)|$, and the phases are shifted in $\pi$
(compare {\sl A} and {\sl B}). The same relation is found between
biphasic-ON and biphasic-OFF filters (compare {\sl C} and {\sl
D}).\normalsize}
\end{center}
\end{figure*}

\begin{eqnarray}\label{equation_9}
   \hat{r}(\omega) = \frac{\sqrt{2\pi}~(1+i~\omega~\tau_{\rm d})}
   {1+i~\omega~\tau_{\rm d} + \sqrt{2\pi}~g~\tau_{\rm
   d}~\hat{h}(\omega)} && \nonumber\\
   \times [ r_0 ~\delta(\omega) &+&
   \hat{h}(\omega)~\hat{s}_1(\omega) ].
\end{eqnarray}

\indent In the absence of feedback ($g$ = 0), this expression
reduces to

\begin{equation}\label{equation_10}
   \hat{r}(\omega) = \sqrt{2\pi}\left[ r_0~\delta(\omega) +
   \hat{h}(\omega)~\hat{s}_1(\omega) \right].
\end{equation}

\indent Solving Eq.~(\ref{equation_10}) for the filter at non-zero
frequencies leads to

\begin{eqnarray} \label{equation_11}
  \hat{h}(\omega) &=& (1/\sqrt{2\pi})~\hat{r}(\omega)/\hat{s}_1(\omega) \nonumber\\
  &=& (1/\sqrt{2\pi})~\hat{r}(\omega)\hat{s}_1^*(\omega) /
  |\hat{s}_1(\omega)|^2,
\end{eqnarray}

\begin{sloppypar}
\noindent so for white-noise stimuli, the filter
in the temporal domain $h(\tau)$ is proportional to the
spike-triggered average \cite{GabbianiKoch1998}.
\end{sloppypar}

\indent When feedback is active ($g \ne 0$), the filter $h(\tau)$
can no longer be calculated with Eq.~(\ref{equation_11}).
Comparing Eqs.~(\ref{equation_9}) and (\ref{equation_10}), we see
that for each frequency $\omega$, the firing probability is still
a linear function of the applied stimulus. The constant of
proportionality, however, is modified due to feedback, and the
modification affects differently the continuous component ($\omega
= 0)$ and the non-zero frequencies ($\omega \ne 0$). For the
continuous case, feedback changes the baseline firing rate $r_0$
to an effective value

\begin{equation}\label{equation_12}
  r_0^{\rm fb} = \frac{h_0 + H \ s_0}{1 + g~\tau_{\rm d}~H} \equiv
  h_0^{\rm fb} + H^{\rm fb} \ s_0,
\end{equation}

\noindent which implements a divisive gain rescaling.

\indent For non-zero frequencies, feedback modifies the coding
properties of the neuron in such a way that the intrinsic filter
$h(\tau)$, processing both the external stimulus and the feedback
signal, is equivalent to an effective filter $h^{\rm fb}(\tau)$
that only filters the external signal. In Fourier space, the
relation between the intrinsic and the effective filters is

\begin{equation}\label{equation_13}
  \hat{h}^{\rm fb}(\omega) = \frac{1+i~\omega~\tau_{\rm d}}
  {1+i~\omega~\tau_{\rm d} + \sqrt{2\pi}~g~\tau_{\rm d}~
  \hat{h}(\omega)} \ \hat{h}(\omega).
\end{equation}

\indent In the remaining part of the present section, we analyze
the effects of Eqs.~(\ref{equation_12}) and (\ref{equation_13})
both in the frequency and the temporal domains, for several types
of filters. We then extend the analysis to imperfect feedback
processes, where the sum in Eq.~(\ref{equation_6}) is no longer
deterministically proportional to $r(t)$. Finally, we consider
non-linear Poisson neurons, and we discuss the validity of the
linear approximation.

\begin{figure*}[t!]
\begin{center}
\includegraphics[scale = 1.00, angle = 0]{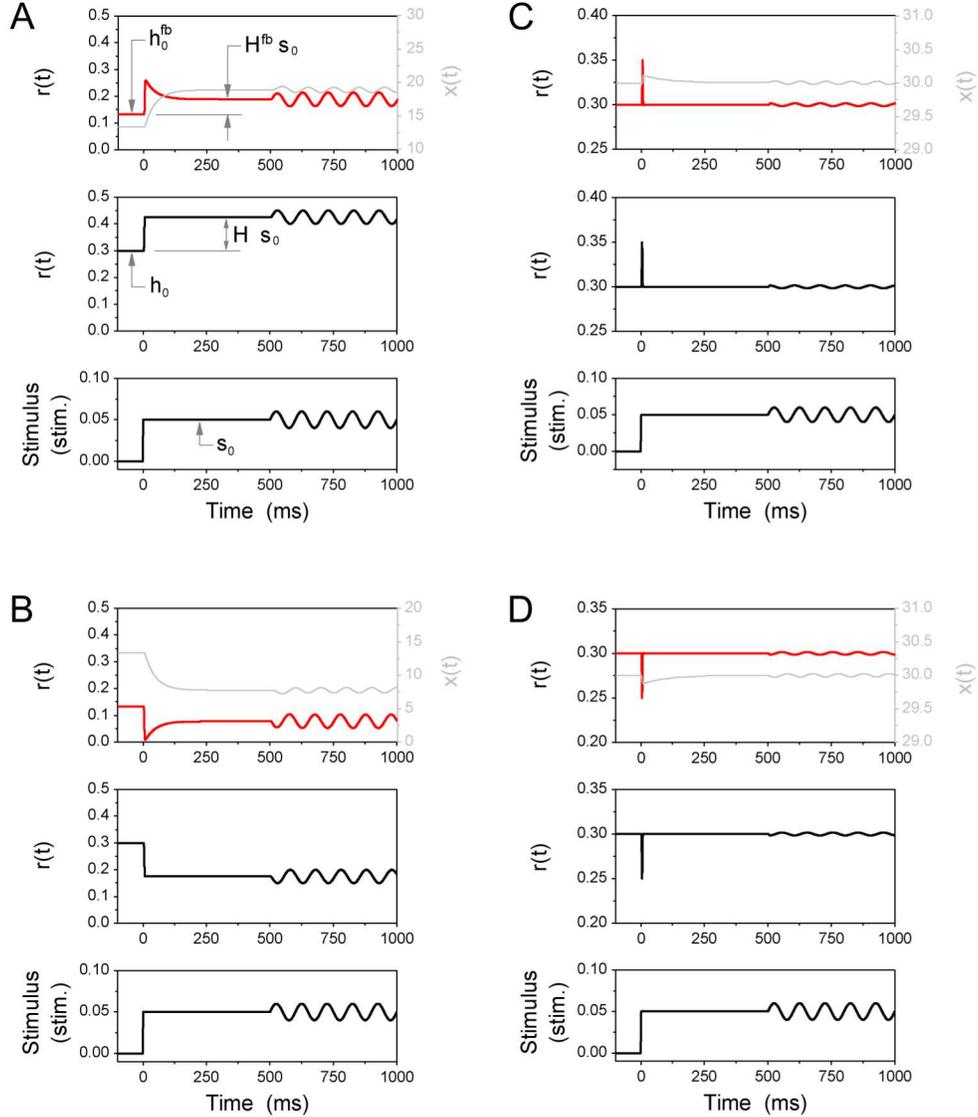}
\caption[Temporal evolution of the firing probability of different
cells, with and without feedback.]{\label{fig2} \linespread{1}
Temporal evolution of the firing probability of different cells,
with and without feedback. {\sl A-D}: ON, OFF, biphasic ON, and
biphasic OFF cell models, respectively (see Fig.~\ref{fig1}).
Bottom panel: Applied stimulus. At time $t=0$~ms, a step of
amplitude $s_{\rm 0} = 0.05$ stimulus units is applied. At time
$t=500$~ms, the sinusoidal stimulation begins (frequency $10$~Hz
and amplitude $0.01$ stimulus units). Intermediate panel: Firing
probability in the absence of feedback. Top panel: Firing
probability $r(t)$ (continuous red line; scale on the left margin)
and feedback process $x(t)$ (continuous gray line; scale on the
right margin) for the model with negative feedback. Shared
parameters: $g = 0.005$ stimulus units, $\tau_{\rm d} = 100$~ms.
Monophasic filters are characterized by $|H| =
2.506~($stim$)^{-1}.($ms$)^{-1}$, whereas biphasic filters are
symmetric, $H =0$. In all cases, $h_{0} = 300$~Hz and the response
is measured in (ms)$^{-1}$. For OFF cells (both monophasic and
biphasic) the coupling factor $g$ is negative.\normalsize}
\end{center}
\end{figure*}

\subsubsection{Poisson neuron models in the absence of feedback}
\indent In order to understand the effect of feedback, we first
describe the basic types of processing in a purely feedforward
model. Typically, neuronal filtering characteristics are
classified according to the shape of $h(\tau)$. For example, in
visual areas, the filters of simple cells are classified in a
limited number of types \cite{Segev_etal2006}: ON, OFF, biphasic
ON and biphasic OFF cells, as shown in Fig.~\ref{fig1}. In a
previous study \cite{UrdapilletaSamengo2009}, we demonstrated that
when processing slow stimuli, the firing probability of these four
different cells is a simple function of the external stimulus. In
ON and OFF cells, $r(t)$ is proportional to $s_1(t - \delta)$,
where the delay $\delta$ is determined by the shape of $h(\tau)$.
In biphasic cells, $r(t)$ is proportional to $s_1'(t - \delta)$,
where $s_1'$ is the temporal derivative of the stimulus. In
Appendix \ref{Apendice2}, we offer a novel derivation of these
results, based on the Fourier approach developed in this paper.

\subsubsection{Poisson neuron models with perfect feedback}
\indent To study the effects of feedback, we separate the analysis
in two: First, we focus on the mean response, determined by the
spectral content of the response at $\omega = 0$. Later, we
analyze non-zero frequencies, $\omega \neq 0$.\\

\indent {\bf Effect of feedback on the baseline firing level}\\
\indent As stated in Eq.~(\ref{equation_12}), the presence of
feedback reduces divisively the firing probability. In
Fig.~\ref{fig2} we observe how the magnitude of the reduction
depends on the type of cell. In the time interval between $0$ and
$500$ ms, the response to a constant stimulus is displayed. Before
the onset of the step stimulus ($t < 0$), the presence of feedback
($g \ne 0$, red curve, upper panels) reduces the steady state
firing probability, as compared to the value obtained in the
absence of feedback ($g = 0$, middle panel), for ON cells
(Fig.~\ref{fig2}{\sl A}). The same holds for the steady-state
firing probability after the step increase in input current, in
the interval between $\sim 250$ and $500$ ms. The stationary
firing level of OFF cells is also diminished (Fig.~\ref{fig2}{\sl
B}) by feedback. In this case, negative feedback is implemented
through a negative coupling factor, $g < 0$, which could, for
example, correspond to excitatory feedback from lateral OFF cells.
An OFF cell with a positive $g$ value could cause the firing
probability to grow unboundedly, and the Fourier transforms to
become ill-defined. For biphasic symmetric cells, the integral of
the filter is zero, $H = 0$, so feedback does not modify the
asymptotic processing of stationary signals (see
Figs.~\ref{fig2}{\sl C} and \ref{fig2}{\sl
D}, for $t<0$).\\

\indent {\bf Effect of feedback on temporal processing}\\
\indent Combining the effective baseline level $r_0^{\rm fb}$ of
Eq.~(\ref{equation_12}) and the effective filter $\hat{h}^{\rm
fb}(\omega)$ defined in Eq.~(\ref{equation_13}), we may now
re-write Eq.~(\ref{equation_9}) as

\begin{equation}\label{equation_14}
  \hat{r}(\omega) = \sqrt{2\pi} \left[r_0^{\rm fb} ~\delta(\omega) +
  \hat{h}^{\rm fb}(\omega)~\hat{s}_1(\omega) \right],
\end{equation}

\noindent which is formally equal to Eq.~(\ref{equation_10}). As a
consequence, in the time domain,

\begin{equation}\label{equation_15}
  r(t) =  r_0^{\rm fb} + \int_{-\infty}^{\infty} h^{\rm fb}(\tau)~s_1(t-\tau)~{\rm
  d}\tau.
\end{equation}

\indent The intrinsic filter $h(\tau)$ characterizes the
biophysical properties of the cell under study. Since $h(\tau)$
filters both the external component of the stimulus $s_0 + s_1(t)$
and the feedback signal $x(t)$, its shape cannot be calculated
through reverse correlation analysis performed with solely the
external signal; knowledge of the internal signal is also
required. The effective filter $h^{\rm fb}(\tau)$, instead, can be
obtained with the sole knowledge of $s_1(t)$. Once $\hat{h}^{\rm
fb}(\omega)$ is obtained, the intrinsic filter $\hat{h}(\omega)$
can be recovered with Eq.~(\ref{equation_13}), assuming that
feedback properties are known. The temporal profile of the
effective filter $h^{\rm fb}(\tau)$ can be obtained simply by
inverse Fourier transform.

\begin{sloppypar}
\indent The Fourier spectrum of the intrinsic filters of both
monophasic and biphasic cells decays at large frequencies (see
Fig.~\ref{fig1}). Hence, in the high frequency range, the
denominator in Eq.~(\ref{equation_13}) is approximately equal to
$1 + i~\omega~\tau_{\rm d}$, and consequently, $\lim_{~\omega \to
\infty} |\hat{h}^{\rm fb}(\omega)| \approx |\hat{h}(\omega)|$.
Biphasic filters also have a reduced spectral content in the low
frequency range (see Figs.~\ref{fig1}{\sl C} and ~\ref{fig1}{\sl
D}). Therefore, in these cells, feedback exerts a limited effect
in the whole frequency range. Since the filtering properties of
these cells are hardly modified by feedback, hereafter we focus on
monophasic cells (ON and OFF).
\end{sloppypar}

\begin{figure}[t!]
\begin{center}
\includegraphics[scale = 1.00, angle = 0]{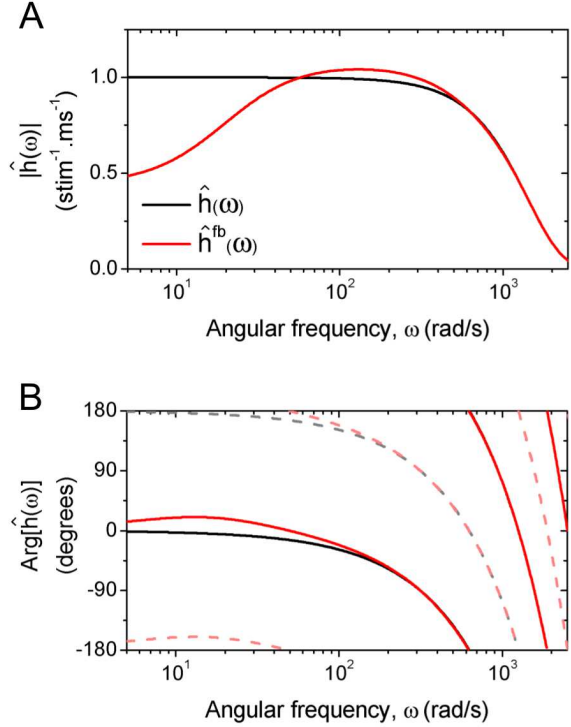}
\caption[Spectral composition of filters of monophasic cells, with
and without feedback.] {\label{fig3} \linespread{1} Spectral
composition of filters of monophasic cells, with and without
feedback ($\hat{h}^{\rm fb}$ in red and $\hat{h}$ in black lines,
respectively). {\sl A}: The modulus $|\hat{h}(\omega)|$ is exactly
the same for ON and OFF cells. Feedback reduces the spectral
content at low frequencies and induces a relative increase in the
intermediate range. {\sl B}: Phase of $\hat{h}^{\rm fb}(\omega)$
and $\hat{h}(\omega)$, for an ON cell (full lines) and an OFF cell
(dashed lines). Feedback advances the phases at intermediate
frequencies. Parameters as in Fig.~\ref{fig2}.\normalsize}
\end{center}
\end{figure}

\indent In Fig.~\ref{fig3}, we discuss the differences in the
spectral domain between $\hat{h}^{\rm fb}(\omega)$ and
$\hat{h}(\omega)$. For any OFF filter there is another ON filter
that is exactly equal in shape, but with inverted sign. To
consider negative feedback, the coupling constant $g$ acting on an
OFF cell also has to be inverted in sign, otherwise, we fall on
the case of positive feedback, which is unstable in the linear
case. Hence, due to inversion symmetry, ON and OFF cells present
exactly the same gain spectral behavior. Additionally, filter
phases of ON and OFF cells display the same spectral
characteristics but shifted in $\pi$.

\begin{figure*}[t!]
\begin{center}
\includegraphics[scale = 1.00, angle = 0]{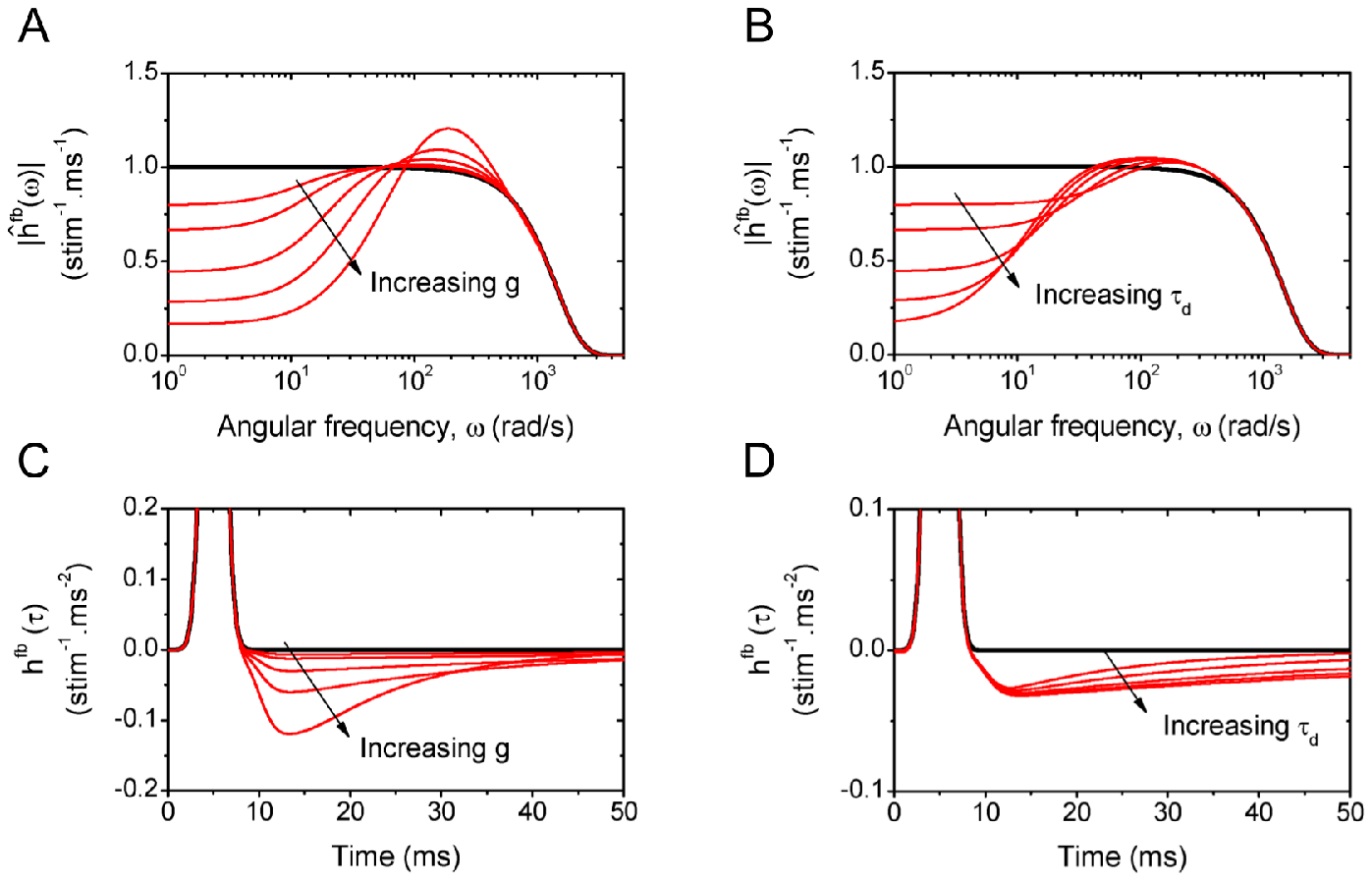}
\caption[Influence of parameters defining feedback on spectral and
temporal processing.] {\label{fig4} \linespread{1} Influence of
parameters defining feedback on spectral and temporal processing.
Black line: intrinsic filter $h$. {\sl A}: Modulus of the feedback
filter $\hat{h}^{\rm fb}(\omega)$, for different values of the
coupling strength $g$ (red lines), with $g =
1,~2,~5,~10$~and~$20~\times 10^{-3}$ stimulus units. For all
cases, $\tau_{\rm d} = 100$~ms. {\sl B}: Modulus of the feedback
filter $\hat{h}^{\rm fb}(\omega)$, for different time constants
$\tau_{\rm d}$ (red lines), with $\tau_{\rm d} = 20,~40,~100,~200
$~and~$400$~ms. For all cases, $g = 5~\times~10^{-3}$ stimulus
units. {\sl C, D}: Temporal filter reconstructed from the spectra
in {\sl A} and {\sl B} and the corresponding phases (not shown).
{\sl C}: As $g$ increases, the region of inverted polarity becomes
more prominent. {\sl D}: As $\tau_{\rm d}$ increases, the recovery
from the inverted polarity becomes slower. In all cases, $H =
2.506~($stim$)^{-1}.($ms$)^{-1}$.\normalsize}
\end{center}
\end{figure*}

\indent Overall, for monophasic cells, we observe a moderate
spectral reshaping of filters due to the presence of feedback. The
low-pass filtering characteristic of $\hat{h}(\omega)$ are
converted to band-pass filtering properties in $\hat{h}^{\rm
fb}(\omega)$ (see black and red continuous lines, respectively, in
Fig.~\ref{fig3}{\sl A}). Feedback reduces the contribution of low
frequency stimuli and slightly enhances the influence of stimuli
with intermediate frequencies. Since the system is linear, the
application of a sinusoidal stimulus (see Fig.~\ref{fig2}, for
$t>500$~ms) evokes a sinusoidal response probability; the ratio
between relative amplitudes of the response and the stimulus
(multiplied by a factor $1/\sqrt{2\pi}$) defines the gain or the
modulus of the filter. The introduction of negative feedback
produces a minor effect on the phase. Importantly, however, a
phase advance is observed at intermediate frequencies, implying
that $r$ overtakes $s$ in the asymptotic regime (see
Fig.~\ref{fig2} for $t \gg 500$~ms).

\indent Feedback is defined by two parameters: the time constant
$\tau_{\rm d}$ that determines the temporal development of $x(t)$,
and its relative contribution to the input signal, given by $g$.
In Figs.~\ref{fig4}{\sl A} and \ref{fig4}{\sl C} we study the
effect of varying the coupling strength $g$ on a monophasic ON
filter. When $g$ is small, feedback barely influences neural
processing. As $g$ increases, low frequency content decreases in
order to satisfy $\hat{h}^{\rm fb}(\omega) \rightarrow H^{\rm fb}
/\sqrt{2\pi} = \hat{h}(0)/(1+g~\tau_{\rm d}~H)$ and band-pass
behavior is emphasized. To see how these characteristics appear in
the temporal domain, we calculate the filter $h^{\rm fb}(\tau)$ by
applying the inverse Fourier transform on $\hat{h}^{\rm
fb}(\omega)$. The effective filter $h^{\rm fb}(\tau)$ is no longer
purely monophasic, as the original filter $h(\tau)$, since it
contains a late phase of inverted polarity (see
Fig.~\ref{fig4}{\sl C}). The absolute value of the integral of the
filter is also reduced, when compared to the original $H$.
Therefore, the feedback filter is more sensitive to the
fluctuations of the stimulus than the original filter
\cite{UrdapilletaSamengo2009}. As the coupling strength $g$ grows,
stronger feedback generates a more significant region of inverted
polarity in $h^{\rm fb}(\tau)$.

\indent The effect of varying $\tau_{\rm d}$ is shown in
Figs.~\ref{fig4}{\sl B} and \ref{fig4}{\sl D}. For fixed $g$, if
$\tau_{\rm d}$ increases, low frequency components decrease
(Fig.~\ref{fig4}{\sl B}). In addition, for small values of
$\tau_{\rm d}$, the gain extends its zero-frequency value, $H^{\rm
fb} / \sqrt{2\pi}$, into a wider range of positive frequencies. As
shown in Fig.~\ref{fig4}{\sl B}, $\tau_{\rm d}$ has only a minor
influence at intermediate frequencies. Finally, in the temporal
domain, the duration of the region of inverted polarity is
governed by $\tau_{\rm d}$ (see Fig.~\ref{fig4}{\sl D}).

\indent To summarize, we conclude that whenever reverse
correlation reveals a filter with biphasic characteristics, the
obtained effective filter $h^{\rm fb}(\tau)$ may not coincide with
the intrinsic filter $h(\tau)$. In particular, at least some of
the biphasic filtering characteristics may derive from negative
feedback.

\subsubsection{Feedback-induced resonances}

\begin{figure}[t!]
\begin{center}
\includegraphics[scale = 1.00, angle = 0]{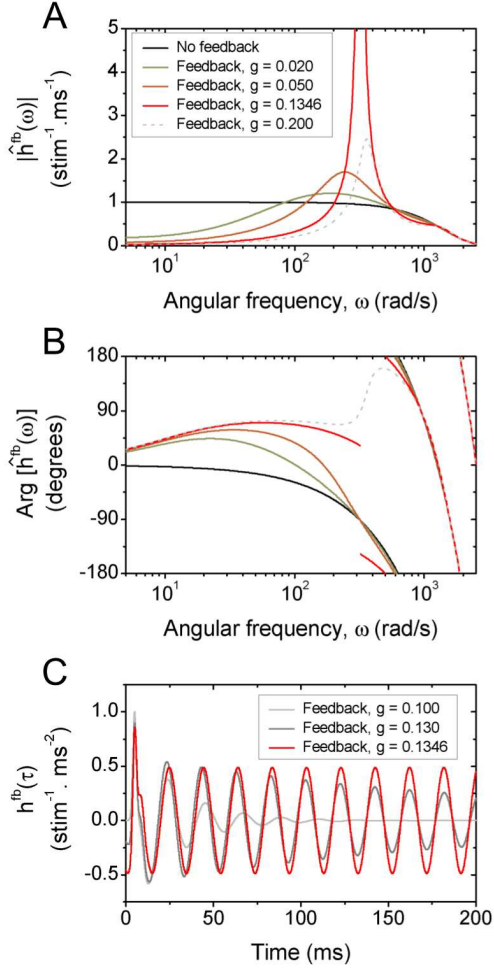}
\caption[Feedback-induced resonances] {\label{fig5} \linespread{1}
Feedback-induced resonances for the ON filter of Fig.~\ref{fig1}.
{\sl A}: Modulus of the effective filter $|\hat{h}^{\rm
fb}(\omega)|$ near the resonant instability. For negative feedback
with timescale $\tau_{\rm d} = 100$~ms, a sharp peak develops at
$\omega \approx 320.4$~rad/s. In this case, the critical value for
the coupling strength is $g_i \approx 0.1346$~stimulus units. {\sl
B}: The phase of the effective filter varies rapidly near the
resonance. {\sl C}: In the temporal domain, as the resonance is
approached, the effective filter develops strong oscillations at
the frequency of the critical $\omega_i$. For $g > g_i$, the
filter $\hat{h}^{\rm fb}(\omega)$ cannot be transformed back to
the time domain, implying that the linear system of
Eq.~(\ref{equation_1}) is ill defined. \normalsize}
\end{center}
\end{figure}

\indent A monophasic receptive field behaves as a low-pass filter,
with a cutoff frequency determined by the inverse of the duration
of the non-zero portion of $h(\tau)$. The effect of feedback is to
additionally reduce the response at low frequencies, so the
effective filter $\hat{h}^{\rm fb}(\omega)$ acquires band-pass
characteristics. The value of the lower cutoff frequency and the
resulting Q-factor depend on the properties of feedback ($g$ and
$\tau_{\rm d}$). It is therefore conceivable that by appropriately
choosing these two parameters, feedback can be shaped as to induce
a strong resonance in the system, even up to the point of
instability. In Eq.~(\ref{equation_13}), this kind of strong
resonance appears as a sharp peak in $\hat{h}^{\rm fb}(\omega)$,
or even a divergence. The denominator of Eq.~(\ref{equation_13})
can indeed vanish for specific combinations of discrete
frequencies $\omega_{j}$ and values of the product $(g \ \tau_{\rm
d})_{j}$. In these cases, the presence of negative feedback
renders the system unstable, amplifying one particular frequency
(or several). Near a resonance, any infinitesimal stimulus
component matching the critical frequency is amplified in the
response, giving rise to strong oscillations. The oscillations
enter repeatedly into the feedback loop producing a divergent
response. Of course, no real neuron can truly produce diverging
responses, because as oscillations grow in amplitude,
Eq.~(\ref{equation_1}) loses validity: The evolution of the system
can no longer be described by a linear equation; in particular,
the linear scheme must be abandoned before the oscillations in the
firing probability $r(t)$ are strong enough as to produce negative
values.

\indent The linear analysis is nevertheless useful to point out
the dramatic amplifying effect that negative feedback can have,
and the conditions that favor resonances. Clearly, if the
denominator of Eq.~(\ref{equation_13}) vanishes for one or more
frequencies, the resulting effective filter $\hat{h}^{\rm
fb}(\omega)$ cannot be transformed back into the temporal domain.
As stated above, such divergences may appear (if at all) at a
discrete collection of frequencies $\omega_{j}$, and discrete
values of the product $(g \ \tau_{\rm d})_{j}$. The absence of
divergences, however, does not guarantee that the inverse Fourier
transforms $h^{\rm fb}(\tau)$ and $r(t)$ exist and are bounded. In
fact, the condition that gives rise to instabilities is broader
and includes the discrete cases where the denominator of
Eq.~(\ref{equation_13}) vanishes. The proper mathematical
framework to analyze the onset of instabilities is provided by
control theory \cite{Franklin_etal1994}, and can be addressed in
terms of the behavior of the Laplace transform of the effective
filter $\tilde{h}^{\rm fb}(s)$ with complex argument $s = \sigma +
i ~ \omega$. As derived in Appendix \ref{Apendice1}, an
instability appears when at least one pole $s_j = \sigma_j + i
~\omega_j$ has positive $\sigma_j$. The pole $s^*$ with largest
real part is hence critical, since the magnitude of $\sigma^*$
determines the stability of the system. When $\sigma^* < 0$, the
effective filter is qualitatively similar to the examples shown in
Figs.~\ref{fig3} and \ref{fig4}. As the coupling strength $g$
increases, $s^*$ gradually shifts to the right and gets closer to
the imaginary axis; consequently, the effective filter begins to
show a prominent peak at the frequency $\omega^*$ of the critical
pole, as shown in Fig.~\ref{fig5}{\sl A}. This resonance is caused
by feedback.

\indent At the resonance, $\omega \approx \omega^*$, the phase of
$\hat{h}^{\rm fb}(\omega)$ varies rapidly (Fig.~\ref{fig5}{\sl
B}). In the temporal domain, the effective filter exhibits strong
oscillations (Fig.~\ref{fig5}{\sl C}), which occasionally grow as
far as to make the system unstable. A mathematical analysis of the
conditions giving rise to instability (see Appendix
\ref{Apendice1}) reveals that unstable behavior is only observed
if $g$ is above a critical threshold, the value of which is
determined by the shape of the intrinsic filter $\hat{h}(\omega)$.
For large $g$, however, the baseline firing level $r_0^{\rm fb}$
drops significantly (see Eq.~(\ref{equation_12})). Therefore, a
transition to instability may only be expected in systems with
strong feedback and, simultaneously, with large intrinsic
spontaneous activity $r_0$ or decreased stimulus fluctuations
$s_1(t)$, so as to ensure that the firing rate $r(t)$ remains
positive. Once these conditions are met, the frequency $\omega^*$
of the unstable oscillations is determined by the time constant
$\tau_{\rm d}$ and the shape of the intrinsic filter $h(\tau)$. In
the limit of large $\tau_{\rm d}$ (as in the example of
Fig.~\ref{fig5}), the feedback time constant becomes irrelevant,
and the location of the critical frequency depends only on
$h(\tau)$.

\subsection{Poisson neuron models with imperfect feedback}

\begin{figure}[t!]
\begin{center}
\includegraphics[scale = 0.6, angle = 0]{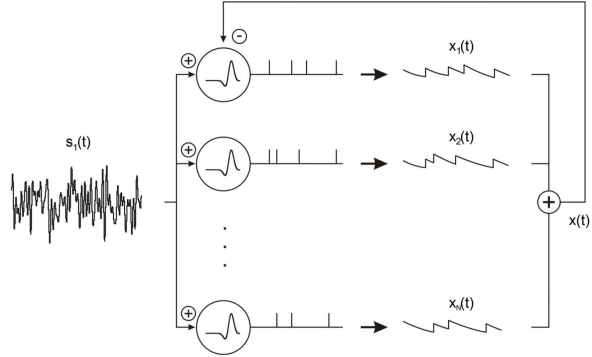}
\caption[Inhibitory feedback supported by a finite population of
neurons.] {\label{fig6} \linespread{1} Inhibitory feedback
supported by a finite population of neurons. The signal $s_1(t)$
stimulates a population of identical neurons. The output spikes
are filtered during synaptic transmission and then return in the
form of a negative feedback input current.\normalsize}
\end{center}
\end{figure}

\indent So far, the analysis was based on Eqs.~(\ref{equation_4})
and (\ref{equation_8}), where the temporal derivative of the
feedback signal $x(t)$ was proportional to the firing probability
$r(t)$. This relation is only valid in the limit of a homogeneous
population of infinitely many identical neurons, uniformly coupled
and driven with identical stimuli $s(t)$. These conditions are
hardly realistic, since the amount of feedback must be determined
by the actual number of generated spikes, and not by the spiking
probability $r(t)$. There is an important difference between these
two options. The spiking probability $r(t)$ is a deterministic
function of the input current $I(t)$ (Eq.~(\ref{equation_1})),
whereas actual spikes are stochastic point processes governed by
$r(t)$. Therefore, in an attempt to provide a more realistic
description of feedback, we now assume that $x(t)$ is given by
Eq.~(\ref{equation_6}). Although this new model can be studied
analytically in specific parameter regimes (see below), we
initially resort to a numerical approach to determine the point up
to which the results of the previous section can be extended to
more realistic conditions.

\indent In the present description, illustrated in
Fig.~\ref{fig6}, we assume that neurons are limited in number, all
have identical intrinsic filtering properties $h(\tau)$, and
process a common temporal stimulus, $s_1(t)$. Furthermore, the
feedback signal $x(t)$ is assumed to be the same for all cells, so
the firing probability $r(t)$ given by Eq.~(\ref{equation_4})
holds for any cell in the population. However, since spike
generation in Poisson processes is stochastic, the precise
temporal location of spikes differs from neuron to neuron and,
therefore, the sum in Eq.~(\ref{equation_6}) is a random variable,
which only recovers its deterministic limit
(Eq.~(\ref{equation_8})) when the number of neurons tends to
infinity ($N \to \infty$). Since the derivative of $x(t)$ is no
longer strictly proportional to $r(t)$, feedback is now called
\textit{imperfect}.

\begin{figure*}[!t]
\begin{center}
\includegraphics[scale = 0.87, angle = 0]{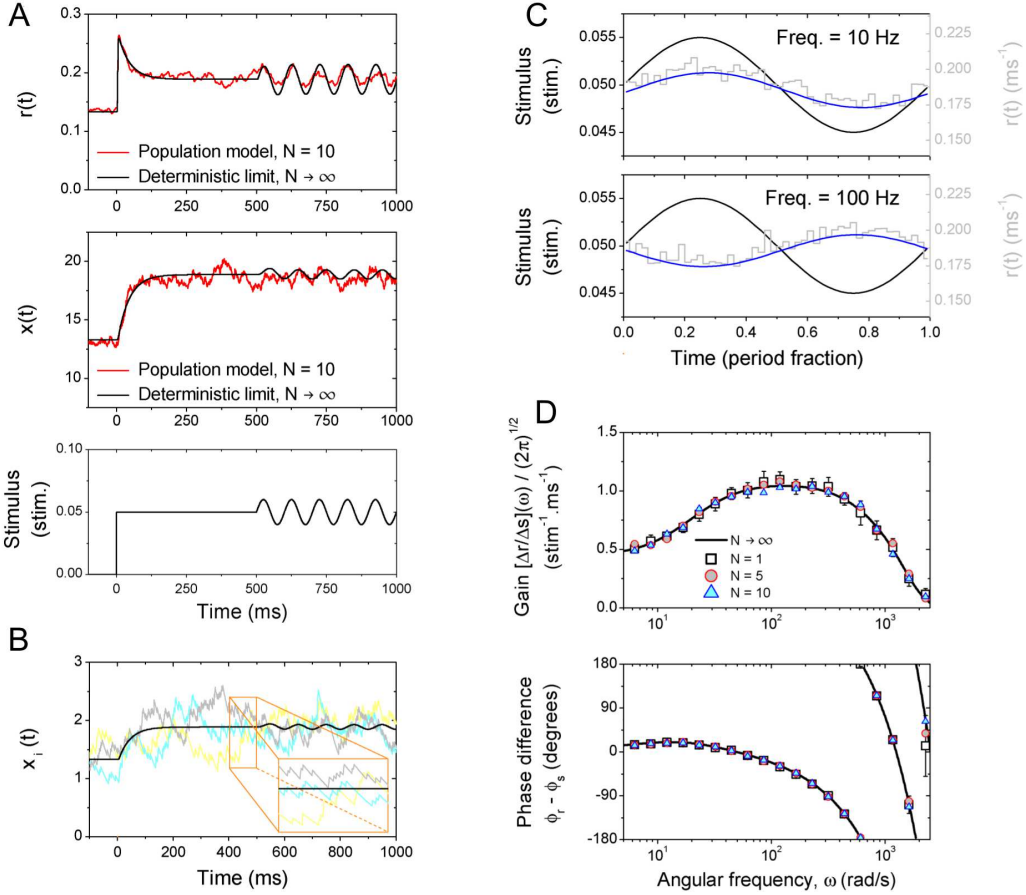}
\caption[Evolution of dynamic variables in the presence of
imperfect feedback] {\label{fig7} \linespread{1} Evolution of
dynamic variables in the presence of imperfect feedback. {\sl A}:
Time-dependent firing probability $r(t)$ and feedback process
$x(t)$ for the model with imperfect feedback driven by the
stimulus $s_0 + s_1(t)$. Black lines: prediction for $N \to
\infty$. Irregular red lines: single-neuron signals, obtained from
a population of $N = 10$ neurons. {\sl B}: Single spike trains
filtered with a simple synaptic dynamics, ${\rm d}x_{\rm i}/ {\rm
d}t = -x_{\rm i}/\tau_{\rm d}+(1/N)\delta(t-t_{\rm i}^{k})$.
Individual signals in colored irregular lines. Irregularity arises
from the discontinuities evoked by incoming spikes, amplified in
the inset. The feedback signal $x(t)$ shown in {\sl A} (middle
panel) is $\sum_{i=1}^{N} x_{\rm i}(t)$ (notice the scale
difference with $x_{\rm i}(t)$ in {\sl B}). Black line:
proportional contribution of a single $x_{\rm i}(t)$ in the limit
$N \to \infty$. {\sl C}: Extracting the relation between stimulus
and response in noisy conditions. For a linear system, a small
sinusoidal stimulus (black line; scale of the left) produces a
sinusoidal firing probability of identical frequency (gray line,
scale on the right). A long run of the output spike train is split
in $T$-windows, where $T$ is the period of the sinusoidal input.
All spikes are then wrapped in a single window, and the histogram
$r(t)$ is constructed. Top/bottom panels: Stimulation at $10~$Hz /
$100~$Hz. {\sl D}: From the fit of the histograms in {\sl C}, we
calculate the gain $\frac{1}{\sqrt{2\pi}} ~\frac{\Delta r}{\Delta
s}$, and the phase, $\phi_{r}-\phi_{s}$, of the transfer function
($\Delta r$ is the amplitude of the sinusoidal function that best
fits the response). Different symbols represent different
population sizes. Spikes are collected in asymptotic conditions
(that is, after the initial transient) during $500$~s. For each
data point, $10$ repetitions are simulated. Error bars indicate
the standard deviation of the gain and phase, as calculated from
the best fit parameters obtained for each of these repetitions.
Parameters: $H = 2.506~($stim$)^{-1}.($ms$)^{-1}$, $\tau_{\rm
d}=100$~ms, $g = 0.005$~stimulus units, $r_{0} = 0.3~$ms$^{-1}$,
$s_0 = 0.05$ and $\Delta s = 0.005$~stimulus units.\normalsize}
\end{center}
\end{figure*}

\indent In Fig.~\ref{fig7}{\sl A} we show the evolution of the
firing probability $r(t)$ (upper panel), along with the feedback
signal $x(t)$ (middle panel) and the driving stimulus (bottom
panel) for the finite population model. These three signals are
common to all neurons. The feedback signal is the sum of $N$
filtered spike trains, a few of which are displayed in
Fig.~\ref{fig7}{\sl B}. The discontinuities in the traces (inset)
are produced by individual spikes. These irregular traces barely
resemble the deterministic counterpart $x_{\rm det}(t)/N$ obtained
in the limit $N \to \infty$ (black curve). Their sum, however,
smoothes fluctuations out, and follows $x_{\rm det}(t)$ closely
(middle panel of Fig.~\ref{fig7}{\sl A}).

\indent To quantify the stochastic behavior of this system, we
drive the cells with a sinusoidal stimulus. As observed in
Fig.~\ref{fig7}{\sl A}, for times $t>500~$ms, the firing
probability has a marked periodic component at the frequency of
the input signal. Therefore, the input/output properties can be
characterized by studying a single stimulus cycle. This reduction
is implemented by taking the spikes fired in different time
windows (one window per stimulus period $T$) and wrapping them
together into a single window, taking care of preserving the
original firing phase with respect to the stimulus. That is, each
spike is displaced an integer number of periods, and located
within a window whose duration is equal to a single period of
stimulation. Once a long spike train realization is so wrapped,
the corresponding histogram in the $T$-window is built, as shown
in Fig.~\ref{fig7}{\sl C}. In this figure, we can clearly observe
the periodic modulation of the response (stairs-like gray lines),
which expectedly can be fitted by a sinusoidal function
(continuous blue lines). The amplitude and phase (relative to the
stimulus) of the adjusted response is used to construct the
spectral characteristics of the stochastic model.

\begin{sloppypar}
\indent In Fig.~\ref{fig7}{\sl C} we show the spectral
characteristics of the population-based feedback model. Both the
gain and the phase shift are practically invariant with the number
of neurons in the population, and a good agreement with the case
$N \to \infty$ (black line) is observed. The dispersion at each
point (error bars) arises from the variability in the sinusoidal
fit, which depends on the irregularities of the histogram
(controlled by the total recording time and the mean number of
spikes produced during a cycle). The irregularities, in turn,
arise from the Poissonian character of the spike generation
process. When $N=1$, the filtered activity of the neuron itself is
used as the feedback signal defined in Eq.~(\ref{equation_6}), and
the sum in $i$ just involves a single element, $i=1$. This
situation is adequate to model self-inhibition due to
spike-triggered adaptation currents \cite{BendaHerz2003,
Benda_etal2010, Urdapilleta2011}. As $N$ grows, the description
gradually shifts to represent network-mediated feedback processes.
\end{sloppypar}

\begin{figure*}[t!]
\begin{center}
\includegraphics[scale = 1.00, angle = 0]{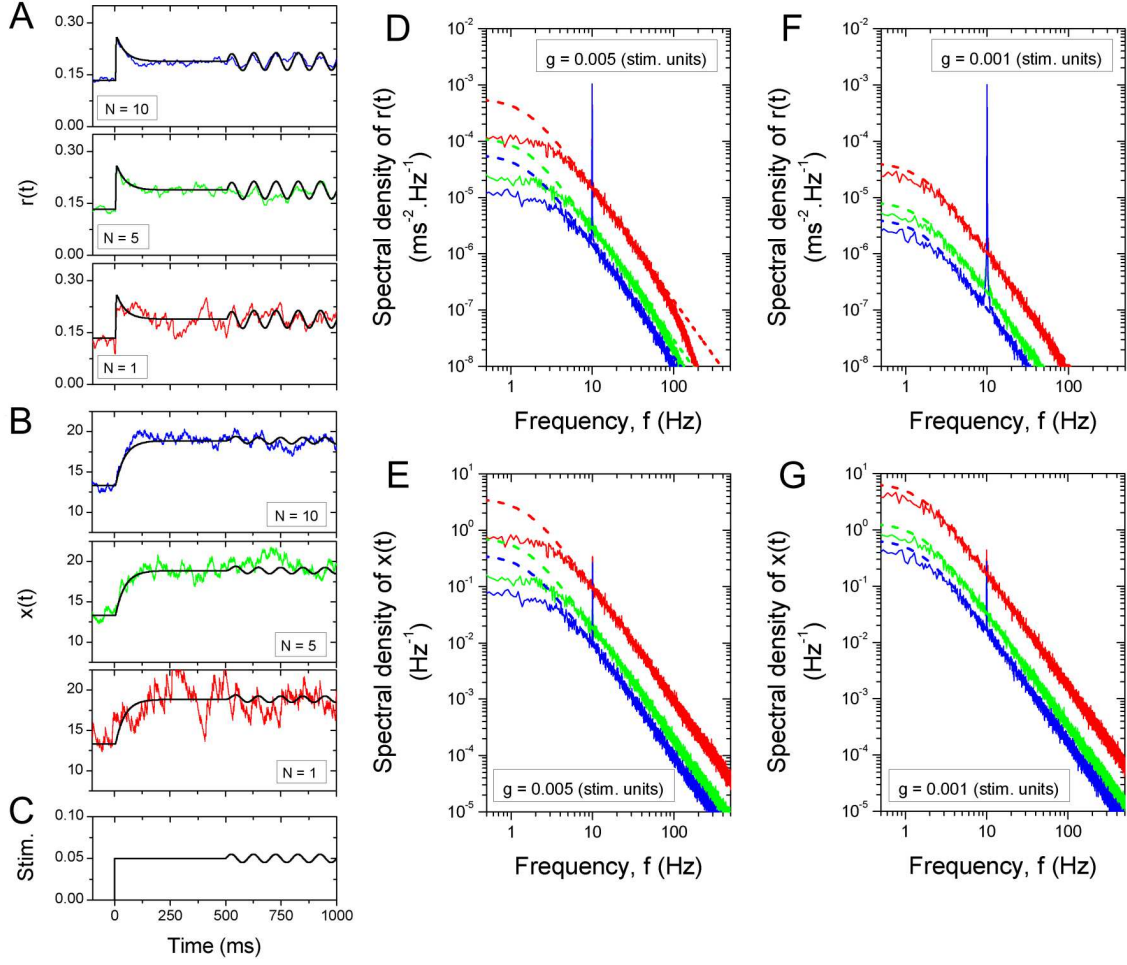}
\caption[Characterization of finite size effects.] {\label{fig8}
\linespread{1} Characterization of finite size effects. {\sl A},
{\sl B}: Evolution of the time-dependent firing rate $r(t)$ (in
ms$^{-1}$) and feedback process $x(t)$ (adimensional), for the
stimulus in {\sl C}. In {\sl A} and {\sl B}, different panels
correspond to different population sizes: $N=10$ (blue, top
panel), $N=5$ (green, middle panel), and $N=1$ (red, bottom
panel). Black lines: deterministic limit $N \to \infty$. All
parameters are the same as those used in Fig.~\ref{fig2}{\sl A}.
As $N$ decreases, fluctuations are amplified. {\sl D, E}: Power
spectral density per unit time of $r(t)$ ({\sl D}) and $x(t)$
({\sl E}), in the asymptotic regime of small-amplitude, periodic
stimulation (frequency: $10~$Hz, amplitude: 0.005 stimulus units).
Colors as in {\sl A} and {\sl B}. The theoretically derived power
spectral density, under the weak coupling assumption, is shown in
dashed lines (with corresponding colors) for $S_{r}(f)$ ({\sl D}),
Eq.~(\ref{equation_25}), and $S_{x}(f)$ ({\sl E}),
Eq.~(\ref{equation_22}). {\sl F, G}: Analogous to panels {\sl D,
E} for a weaker feedback, $g =0.001$~stimulus units.\normalsize}
\end{center}
\end{figure*}

\indent The wrapping procedure used to construct Fig.~\ref{fig7}
allows us to determine the linear response function as the ratio
between the amplitude of the firing probability (response) and the
stimulus driving the system (input), both measured at the same
frequency. One important result is that the linear response
function is independent of the population size $N$ (see
Fig.~\ref{fig7}{\sl D}). However, the irregularities observed in
the temporal response do indeed depend on $N$. To exemplify this
behavior, in Figs.~\ref{fig8}{\sl A} and \ref{fig8}{\sl B} we show
the time-dependent firing probability $r(t)$ and the feedback
process $x(t)$, for the cases $N = 10$, $N = 5$ and $N = 1$.
Irregularities become more prominent as the population size
decreases. When the number of neurons is finite, the firing
probability contains a certain amount of power at frequencies
different from the incident frequency. The power at spurious
frequencies does not affect the linear response function, because
the fitting procedure is casted specifically at the input
frequency. So far we have described the properties of the response
at this input frequency. Now we turn our attention to the rest of
the spectrum.

\indent As an example, in Figs.~\ref{fig8}{\sl D} and
\ref{fig8}{\sl E} we show the power spectral density of $r(t)$ and
of $x(t)$, when the system is stimulated with a $10$~Hz,
small-amplitude sinusoidal signal. Given the linearity of the
system, the firing probability and the feedback signal contain a
strong component at precisely $10$~Hz (see the peaks in the
corresponding spectra). The height of the peak in $r(t)$ is given
by the linear response function studied before (see
Fig.~\ref{fig7}). The remaining spectral power (the background)
arises from the inherent randomness of Poisson processes. As the
population size increases, fluctuations in the spontaneous regime
diminish and, correspondingly, the background spectral density
decreases as well.

\indent  The remaining part of this section is devoted to obtain
an analytical expression of the background spectrum. We first
analyze the feedback process corresponding to $N=1$ (adaptation
current), and later generalize the result to arbitrary $N$. For $N
= 1$, the signal $x(t)$ evolves according to

\begin{equation}\label{equation_16}
   \frac{{\rm d}x}{{\rm d}t} = -\frac{x}{\tau_{\rm d}} +
   \xi(t),
\end{equation}

\noindent where $\xi(t)$ represents the spike train produced by a
Poisson process (a realization), with a time-dependent rate given
by Eq.~(\ref{equation_4}). This rate is coupled to
Eq.~(\ref{equation_16}) (and thereby, to the noise source) through
the feedback term. By approximating the time-dependent firing rate
by its baseline level $r_0^{\rm fb}$ (Eq.~(\ref{equation_12})),
Eq.~(\ref{equation_16}) becomes a standard linearly filtered
Poisson process (filtered \textit{shot} noise). Under this
approximation, it is simple to find the formal solution to
Eq.~(\ref{equation_16}) and the resulting exponential
autocorrelation function,

\begin{equation}\label{equation_17}
   C_{x}(\tau) = \frac{1}{2}~r_0^{\rm fb}~\tau_{\rm d}~
   {\rm e}^{-|\tau|/\tau_{\rm d}}.
\end{equation}

\indent The power spectral density (one-sided, per unit time) of
the feedback signal $x(t)$ can be obtained from the
autocorrelation function through the Wiener-Khinchin theorem
\cite{Gardiner1985, Press_etal2007},

\begin{eqnarray}\label{equation_18}
   \frac{S_{x} (\omega)}{T} &=& \frac{1}{T} \left[ |\hat{x}(-\omega)|^{2}
   + |\hat{x}(\omega)|^{2} \right] \nonumber\\
   &=& \frac{1}{\sqrt{2\pi}}~\left[
   \hat{C}_{x}(-\omega) + \hat{C}_{x}(\omega) \right].
\end{eqnarray}

\indent Since the Fourier transform of Eq.~(\ref{equation_17})
reads

\begin{equation}\label{equation_19}
   \hat{C}_{x}(\omega) = \hat{C}_{x}(-\omega) = \frac{r_0^{\rm fb}}
   {\sqrt{2\pi}}~\frac{1}{(1/\tau_{\rm d})^2 + \omega^{2}},
\end{equation}

\noindent the one-sided power spectral density per unit time is

\begin{equation}\label{equation_20}
   \frac{S_{x} (\omega)}{T} = \frac{r_0^{\rm fb}}{\pi}~
   \frac{1}{(1/\tau_{\rm d})^2 + \omega^{2}},
\end{equation}

\noindent or, in terms of frequency,

\begin{equation}\label{equation_21}
   \frac{S_{x} (f)}{T} = 2\pi~\frac{S_{x} (\omega)}{T} =
   \frac{2~r_0^{\rm fb}}{(1/\tau_{\rm d})^2 + (2\pi f)^{2}}.
\end{equation}

\begin{sloppypar}
\indent The previous analysis can be easily
extended to the case $N > 1$. In this case, the effective rate for
$\xi(t)$ in Eq.~(\ref{equation_16}) is now $N~r_0^{\rm fb}$. We
recall that, in order to maintain the feedback level constant when
$N$ increases, the efficacy of each spike in
Eq.~(\ref{equation_6}) has to be proportional to $1/N$. Therefore,
now ${\rm d}x/{\rm d}t = -x/\tau_{\rm d} + (1/N)~\xi(t)$, and the
autocorrelation function is a scaled version of
Eq.~(\ref{equation_17}), $C_{x}^{N\,{\rm neurons}}(\tau) =
C_{x}^{1\,{\rm neuron}}(\tau)/N$. Consequently,
\end{sloppypar}

\begin{equation}\label{equation_22}
   \frac{S_{x} (f)}{T} = \frac{1}{N}~
   \frac{2~r_{\rm 0}^{\rm fb}}{(1/\tau_{\rm d})^2 + (2\pi f)^{2}}.
\end{equation}

\indent This expression is represented with dashed lines in
Fig.~\ref{fig8}{\sl E}, for different population sizes. The power
spectrum obtained from simulations agrees with the theoretical
description, except at low frequencies, where higher order
statistical interactions between $x(t)$ and $r(t)$ become
noticeable. At medium and high frequencies, the theoretical
approach provides a very good description of the simulations.

\indent The firing probability $r(t)$ inherits the correlation
structure of $x(t)$,

\begin{eqnarray}
   C_{r}(\tau) &=& g^{2}~ \int_{0}^{\tau_{\rm m}} h(\tau')~{\rm
   d}\tau'~\int_{0}^{\tau_{\rm m}} h(\tau'')~C_{x}^{(N)}(\tau+\tau'-\tau'')
   ~{\rm d}\tau'',\nonumber\\
   \label{equation_23}
   &=& \frac{1}{2N}~g^{2}~r_{\rm 0}^{\rm fb}~\tau_{\rm d} \int_{0}^{\tau_{\rm m}} h(\tau')~{\rm
   d}\tau'~\int_{0}^{\tau_{\rm m}} h(\tau'')\nonumber\\
   &&\hspace{3cm}\times~{\rm e}^{-|\tau+\tau'-\tau''|/\tau_{\rm d}}~
   {\rm d}\tau''.
\end{eqnarray}

\indent The filter $h(\tau)$ is different from zero inside a
finite window $[0,\tau_{\rm m}]$ \cite{UrdapilletaSamengo2009}.
Therefore, in Eq.~(\ref{equation_23}), we replaced the upper
integration limits by $\tau_{\rm m}$. Whenever $\tau_{\rm m} \ll
\tau_{\rm d}$, the exponential factor in the integrand of
Eq.~(\ref{equation_23}) can be further simplified, so that only
time differences $\tau$ comparable or larger than $\tau_{\rm m}$
matter:

\begin{itemize}
\item[\textbullet] $\tau \sim \mathcal{O}(\tau_{\rm
m})~~\Rightarrow~~{\rm e}^{-|\tau+\tau'-\tau''|/\tau_{\rm d}}
\approx 1$, \item[\textbullet] $\tau \gg \mathcal{O}(\tau_{\rm m})
~~\Rightarrow~~{\rm e}^{-|\tau+\tau'-\tau''|/\tau_{\rm d}} \approx
{\rm e}^{-|\tau|/\tau_{\rm d}}$.
\end{itemize}

\indent In this case, the autocorrelation function can be
approximated by

\begin{equation}\label{equation_24}
   C_{r}(\tau) = \frac{1}{2N}~g^{2}~H^{2}~r_{\rm 0}^{\rm fb}~\tau_{\rm
   d}~{\rm e}^{-|\tau|/\tau_{\rm d}}.
\end{equation}

\indent Finally, based on this equation and the Wiener-Khinchin
theorem, the power spectral density for the firing probability
$r(t)$ is

\begin{equation}\label{equation_25}
   S_{r} (f) = \frac{2}{N} \ \frac{~g^{2}~H^{2}~
   r_0^{\rm fb}~}{(1/\tau_{\rm d})^2 + (2\pi f)^{2}}.
\end{equation}

\indent This expression is represented in Fig.~\ref{fig8}{\sl D}
for different population sizes, in dashed lines. As in the
previous analysis, at low frequencies, the coupling between $x(t)$
and $r(t)$ produces a small discrepancy between the numerical
results and the theoretical expression. At high frequencies, a
faster decay than the predicted $S_{r}(f) \sim 1/f^{2}$ is
observed, originated by the approximation at $\tau \sim
\mathcal{O}(\tau_{\rm m})$ during the assessment of the
autocorrelation function $C_{r}(\tau)$.

\indent The derivation in this section is based on the hypothesis
that the firing rate $r(t)$ could be approximated by its baseline
level $r_{\rm 0}^{\rm fb}$. This assumption is valid if $s_1(t)$
is small, and if $x(t)$ and $r(t)$ are weakly coupled. Obviously,
as the coupling strength $g$ becomes smaller, the theoretical
expressions become more accurate, and are valid in a wider range
of frequencies. As shown in Figs.~\ref{fig8}{\sl F} and
\ref{fig8}{\sl G}, in this limit, the simulated spectral densities
are in excellent agreement with Eqs.~(\ref{equation_22}) and
(\ref{equation_25}).

\subsection{Linear-nonlinear Poisson neuron models}
\begin{sloppypar}
\indent Linear Poisson models are only an approximate description
of the processes governing neuronal dynamics. We here improve the
approximation by adding a static nonlinearity, as often done in
the description of sensory systems \cite{DayanAbbott2001,
Chichilnisky2001, BaccusMeister2002, GaudryReinagel2007,
Sharpee_etal2008, Butts_etal2011, GarvertGollisch2013}. The new
model constitutes the linear-nonlinear Poisson approach, where
\end{sloppypar}

\begin{equation}\label{equation_26}
   r(t) =  f \left[ r_0 + \int_{-\infty}^{\infty} h(\tau)~[s_1(t-\tau) -
   g~x(t-\tau)]~{\rm d}\tau \right].
\end{equation}

\indent In this section we demonstrate that although the results
of the previous section do not strictly hold in the presence of a
nonlinearity, it is possible to develop an approximate version of
the theory that makes very good predictions in most practical
cases.

\begin{figure*}[t!]
\begin{center}
\includegraphics[scale = 0.9, angle = 0]{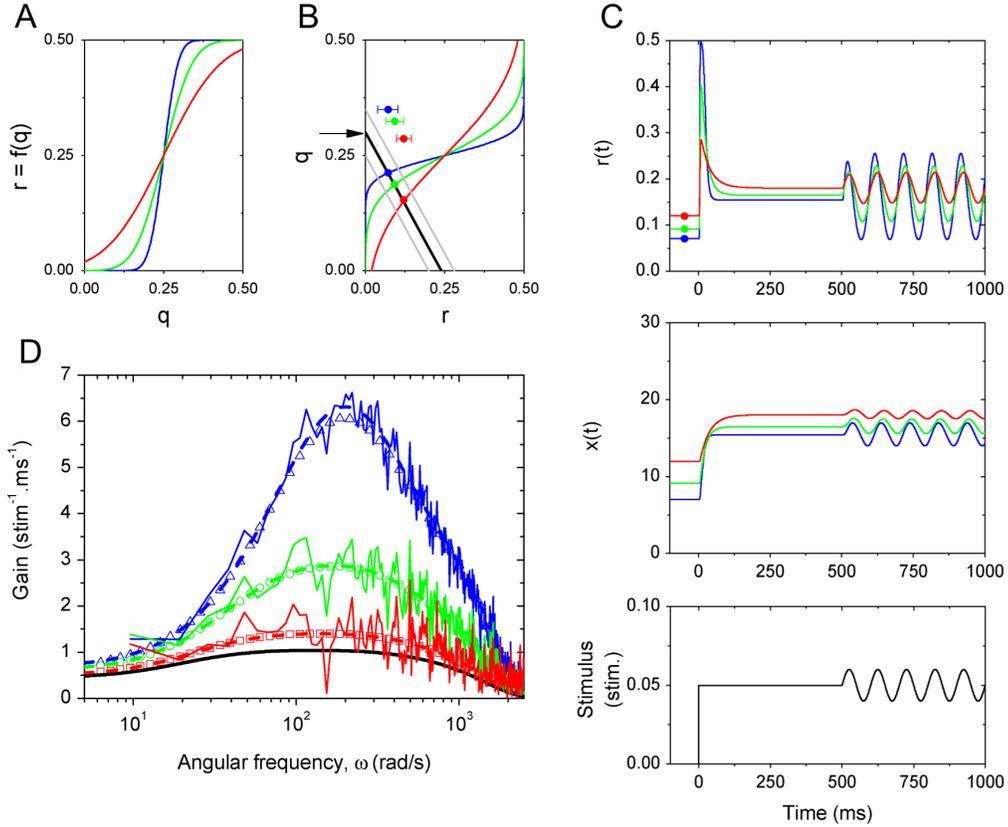}
\caption[Effects of a static nonlinearity on the feedback model.]
{\label{fig9} \linespread{1} Effects of a static nonlinearity on
the feedback model. {\sl A}: Sigmoidal nonlinearities modeled as
\textit{error functions}, $(1/2)~r_{\rm max}~\{ {\rm
erf}[(q-q_{\rm c})/\Delta] + 1\}$, with $\Delta = 0.05$, $\Delta =
0.10$, and $\Delta = 0.20$ (blue, green, and red lines,
respectively). In all cases, $q_{\rm c} = 0.25$ and $r_{\rm max} =
0.50$. Axis units in ms$^{-1}$. {\sl B}: The spontaneous firing
rate is defined by the intersection between the line $(r_0 -
g~\tau_{\rm d}~H~r)$ (thick black line) and the inverse function
of the nonlinearity (colored curves in {\sl A}). Spontaneous
firing rates for the different nonlinearities are indicated by
colored circles, for $h_{0} = 0.30$ ms$^{-1}$. The addition of a
static positive or negative stimulus (gray lines) shifts the
operation point nonlinearly (circles plus error bars in the upper
part of the figure). {\sl C}: Temporal evolution of the firing
probability (top panel) and the feedback process (middle panel),
under the effect of a particular time-dependent stimulus (bottom
panel). In the absence of stimulus, $t<0$~ms, spontaneous firing
rates are defined by the colored circles in {\sl B}. The step
stimulus deflects the firing probability transiently, until a new
stationary value is reached. A periodic stimulus, $t>500$~ms,
evokes a periodic response probability, whose amplitude is a
non-linear function of the stimulus' amplitude. {\sl D}: Spectral
composition of filters' gain in the non-linear model. Continuous
black line: gain of the feedback filter for the linear model, see
Fig.~\ref{fig3}{\sl A}. Colors match the non-linearities shown in
{\sl A}. Symbols: ratio between the amplitudes of the sinusoidal
response and stimulus, for a small sinusoidal stimulation, as a
function of the driving frequency, $(1/\sqrt{2\pi})\Delta r /
\Delta s$. Irregular lines: gains of the Fourier transform of the
filters obtained from the spike-triggered average of the nonlinear
model, equivalent to $\hat{\chi}^{\rm fb}(\omega)$ for small
stimuli, see Eq.~(\ref{equation_35}). Dashed lines: equivalent
linear filter of the non-linear model $\hat{h}^{\rm nl,
fb}(\omega)$, which can be approximated by an appropriate scaling
of the linear filter, see Eq.~(\ref{equation_37}). Parameters are
identical to those used before for the linear model (see
Figs.~\ref{fig2} and \ref{fig3}).\normalsize}
\end{center}
\end{figure*}

\indent In Fig.~\ref{fig9}{\sl A}, we show three sigmoidal
nonlinearities $f$, all with the same functional shape, but with
different scaling parameters in the $x$-axis. Here we show how to
produce an approximate linearized model, and discuss its range of
validity. The first step is to find the operation point, that is,
the spontaneous firing rate $r_0^{\rm nl}$, that may differ from
the spontaneous rate of the linear case $r_0^{\rm fb}$
(Eq.~(\ref{equation_12})). In the absence of time-dependent
external stimuli, $s_1(t) = 0$, the linear-nonlinear model reduces
to

\begin{equation}\label{equation_27}
   r_0^{\rm nl}  =  f \left[ r_0 - g~\tau_{\rm d}~H~r_0^{\rm nl}\right]
   \Rightarrow f^{-1}(r_0^{\rm nl})  =  r_0 - g~\tau_{\rm d}~H~r_0^{\rm nl}.
\end{equation}

\begin{sloppypar}
\indent Equation~(\ref{equation_27}) implicitly defines $r_0^{\rm
nl}$ as the intersection between the function $f^{-1}(r_0^{\rm
nl})$ and the straight line $r_0 - g~\tau_{\rm d}~H~r_0^{\rm nl}$.
In Fig.~\ref{fig9}{\sl B}, we illustrate the procedure. The level
$r_0$ (arrow on the left margin) sets the offset of the straight
line (thick black line). The slope of the line is determined by
the intrinsic properties of the neuron ($H$) and feedback ($g \
\tau_{\rm d}$). Depending on the steepness of the nonlinearity,
one same $r_0$ may elicit different spontaneous firing rates
$r_0^{\rm nl}$ (colored circles located at the intersections).

\indent If the constant stimulus component $s_0$ is modified, a
new value $r_0$ is established and, after a brief transient
evolution, the operation level $r_0^{\rm nl}$ sets to a new value,
as predicted by Eq.~(\ref{equation_27}). The $y$-intercept of the
straight line shifts, thus displacing the stationary firing
probability. In Fig.~\ref{fig9}{\sl B} the shift is represented by
the two parallel lines (thin gray lines), corresponding to the
addition of two constant stimuli of opposite signs. Due to the
non-linear nature of the model, the same stimuli displace the
response by different amounts (lengths of the error bars).
Moreover, positive and negative stimuli produce effects of
different magnitude (compare the lengths of the left and the right
portions of the error bars).
\end{sloppypar}

\indent In Fig.~\ref{fig9}{\sl C} the temporal evolution of the
firing rate (top panel) and the feedback process (middle panel) is
shown, for a temporally complex input signal. In the spontaneous
regime, $t<0$~ms, the firing probabilities obtained in
Fig.~\ref{fig9}{\sl B} are indicated with colored circles.
Following the application of a step stimulus, the firing
probability and the feedback process undergo a transient evolution
that rapidly settles onto a new stationary value. When the
curvature of the nonlinearity is mild, the evolution is similar to
the linear case (compare the evolution of the red line in
Fig.~\ref{fig9}{\sl C} with the one of Fig.~\ref{fig2}{\sl A}). As
the nonlinearity becomes steeper (blue line in Fig.~\ref{fig9}{\sl
C}), the firing probability evolves faster, and the initial
transient becomes stronger.

\indent The next step is to determine how time-dependent stimuli
are processed. When a sinusoidal signal is applied ($t>500$~ms, in
Fig.~\ref{fig9}{\sl C}), the response probability is periodic, but
not necessarily sinusoidal. Only in the limit of small $s(t)$ is
the sinusoidal response guaranteed, and in this limit, the ratio
between the (mean-subtracted) relative amplitudes of the
input/output sinusoidal modulations is independent of the
amplitude of the stimulus. The amplitude of the response, however,
depends on the steepness of the nonlinearity (Fig.~\ref{fig9}{\sl
D}).

\indent To understand the spectral processing of the non-linear
model, we define the transformed firing rate

\begin{equation}\label{equation_28}
   q(t) = f^{-1}[r(t)],
\end{equation}

\noindent so that Eq.~(\ref{equation_26}) becomes

\begin{equation}\label{equation_29}
   q(t) =  r_0 + \int_{-\infty}^{\infty} h(\tau)~[s_1(t-\tau) -
   g~x(t-\tau)]~{\rm d}\tau.
\end{equation}

\indent In terms of $q(t)$, the feedback current $x(t)$,
previously described by Eq.~(\ref{equation_8}), is now governed by

\begin{equation} \label{equation_30}
   \frac{{\rm d}x}{{\rm d}t} = - \frac{x}{\tau_{\rm d}} + f\left[q(t)\right].
\end{equation}

\indent Equations (\ref{equation_29}) and (\ref{equation_30})
constitute a closed system, but the presence of the nonlinearity
precludes the application of the linear Fourier approach that
allowed us, in the previous section, to find the relation between
$h(\tau)$ and $h^{\rm fb}(\tau)$. Linearizing
Eq.~(\ref{equation_30}) around the operation point, $q_0 =
f^{-1}(r_0^{\rm nl})$, yields

\begin{equation} \label{equation_31}
   \frac{{\rm d}x}{{\rm d}t} \approx - \frac{x}{\tau_{\rm d}} +
   r_0^{\rm nl} + f'(q_0) \, \left[ q(t)-q_0 \right].
\end{equation}

\indent Equations (\ref{equation_29}) and (\ref{equation_31}) are
closed and linear, and therefore, allow for a linear treatment.
Transforming them both to Fourier space, and after some algebraic
manipulations,

\begin{eqnarray}\label{equation_32}
   \hat{q}(\omega) &=& \sqrt{2\pi} \Bigg[ q_0 \,
   \delta(\omega) \nonumber\\
   && + \frac{(1+i~\omega~\tau_{\rm d})\,\hat{h}(\omega)}{1+i~\omega~
   \tau_{\rm d} + \sqrt{2\pi}~g~\tau_{\rm d}~f'(q_0)
   ~\hat{h}(\omega)} \, \hat{s}_1(\omega) \Bigg].
\end{eqnarray}

\indent Equation~(\ref{equation_32}) can be written in terms of an
effective filter $\hat{\chi}^{\rm fb}(\omega)$, such that

\begin{equation}\label{equation_33}
   \hat{q}(\omega) = \sqrt{2 \pi} \left[q_0 \, \delta(\omega)
   + \hat{\chi}^{\rm fb}(\omega) \, \hat{s}_1(\omega) \right],
\end{equation}

\noindent where

\begin{equation}\label{equation_34}
   \hat{\chi}^{\rm fb}(\omega) = \frac{(1+i~\omega~\tau_{\rm d})
   \,\hat{h}(\omega)}{1+i~\omega~\tau_{\rm d} + \sqrt{2\pi}~g~
   \tau_{\rm d}~f'(q_0)~\hat{h}(\omega)}.
\end{equation}

\indent The possibility of summarizing the effect of feedback in
Eq.~(\ref{equation_33}) implies that in the temporal domain,

\begin{equation} \label{equation_35}
  r(t) = f \left[q_0 + \int_{-\infty}^{+\infty} \chi^{\rm fb}(\tau) \ s_1(t -
  \tau) \ {\rm d}\tau \right].
\end{equation}

\begin{sloppypar}
\indent The presence of the nonlinearity implies that the
effective filter $\hat{\chi}^{\rm fb}(\omega)$ can no longer be
calculated as the ratio between $\hat{r}(\omega)$ and
$\hat{s}_1(\omega)$. Reverse correlation, however, still allow us
to calculate $\chi^{\rm fb}(\tau)$ from the spike triggered
average of the recorded data \cite{GabbianiKoch1998,
Chichilnisky2001}. The result is illustrated by the irregular
lines in Fig.~\ref{fig9}{\sl D}.
\end{sloppypar}

\indent Reverse correlation provides the best possible estimate of
$\hat{\chi}^{\rm fb}(\omega)$, since the approach only entails the
linearization of Eq.~(\ref{equation_8}). The method, however,
requires large amounts of data to converge to a reliable
estimation ($2.5\times 10^{6}$ spikes, in Fig.~\ref{fig9}{\sl D}).
For practical purposes, hence, one may be willing to sacrifice
some modeling accuracy, and further linearize
Eq.~(\ref{equation_35}), for the sake of obtaining an easier
estimation method. Such approximation brings the response and the
stimulus to be linearly related

\begin{equation} \label{equation_36}
  r(t) \approx f(q_0) + f'(q_0) \, \int_{-\infty}^{+\infty}
  \chi^{\rm fb}(\tau) \ s_1(t - \tau) \ {\rm d}\tau,
\end{equation}

\noindent so that the linear response function of the non-linear
system is

\begin{equation}\label{equation_37}
   \hat{h}^{\rm nl, fb}(\omega) = \frac{(1+i~\omega~\tau_{\rm d})\,f'(q_0)}
   {1+i~\omega~\tau_{\rm d} + \sqrt{2\pi}~g~
   \tau_{\rm d}~f'(q_0)~\hat{h}(\omega)}\,\hat{h}(\omega),
\end{equation}

\noindent differing from the linear approach
(Eq.~(\ref{equation_13})) by a rescaling controlled by the gain at
the operation point.

\indent  In Fig.~\ref{fig9}{\sl D}, the spectral behavior of the
filter obtained from the spike-triggered average of the non-linear
model, $\hat{\chi}^{\rm fb}(\omega)$, as well as the linearized
filter $\hat{h}^{\rm nl, fb}(\omega)$ are shown. The symbols
represent the results obtained from the ratio between
$\hat{r}(\omega)$ and $\hat{s}_1(\omega)$, for small sinusoidal
stimuli (defined as in Fig~\ref{fig7}{\sl D}). As observed, both
filters are in excellent agreement with the simulated data.
Clearly, as the non-linearity becomes steeper, the band-pass
characteristics are more pronounced (see also, temporal responses
in Fig.~\ref{fig9}{\sl C}).

\begin{figure*}[t!]
\begin{center}
\includegraphics[scale = 1.00, angle = 0]{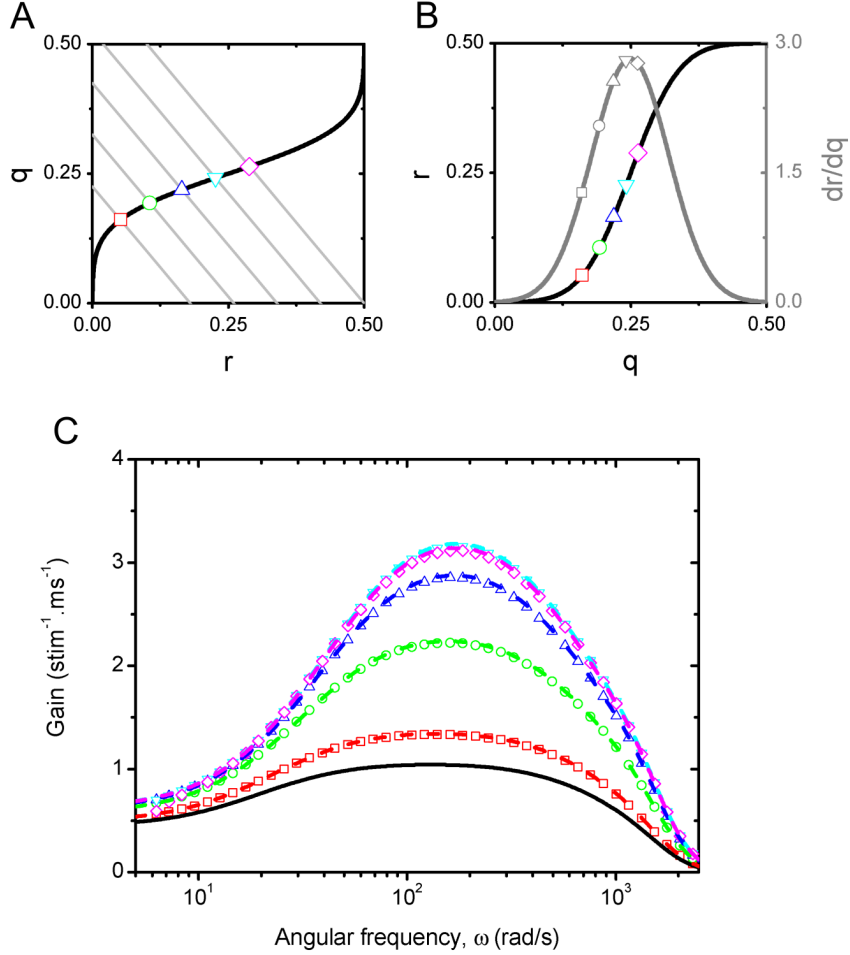}
\caption[Influence of the operation point in the nonlinear
feedback model.] {\label{fig10} \linespread{1} Influence of the
operation point in the nonlinear feedback model. {\sl A}: A
sigmoidal nonlinearity operated at different points. Different
straight lines correspond to different values of $h_{0}$ ($0.1$,
$0.2$, $0.3$, $0.4$ y $0.5~$ms$^{-1}$, respectively). The
$y$-intercept is $h_{0}+H~s_{\rm 0}$, with $H =
2.506~($stim$)^{-1}.($ms$)^{-1}$ and $s_{\rm 0} = 0.05$ stimulus
units, and the nonlinearity is defined as in Fig.~\ref{fig9}{\sl
A}, with $\Delta = 0.10$. {\sl B}: Firing probabilities, for the
different values of $h_{0}$ defined in {\sl A}, indicated by
symbols located on the nonlinearity (thick black line, scale on
the left margin). Associated gains are represented with
corresponding symbols on the derivative of the function (gray
line, scale on the right margin). {\sl C}: Ratio between relative
amplitudes of response and stimulus signals, for different driving
frequencies. Symbols and colors correspond to those represented in
{\sl A}. Black thick line displays the spectral behavior of the
lineal model, whereas colored dashed lines are appropriately
scaled versions.\normalsize}
\end{center}
\end{figure*}

\indent The processing differences so far described for
nonlinearities of varying steepness are also observed in a single
nonlinearity, at varying operation points. In Fig.~\ref{fig10}{\sl
A}, the operation point is varied by manipulating the value of
$h_{0}$. The stationary firing rates are represented in
Fig.~\ref{fig10}{\sl B} by symbols located on the nonlinearity
(thick black line), whereas the gains are indicated on the
derivative (gray line). The comparison between the nonlinear model
with the linear approximations is similar to the comparison of
Fig.~\ref{fig9}{\sl D}.

\indent In the purely linear case, we observed that in certain
conditions, feedback could give rise to unstable dynamics.
Resonances, understood as peaks in the power spectrum of the
effective filter, are still possible in the nonlinear case.
However, now the resonant peak cannot grow indefinitely, so
unstable divergences are ruled out. The linear approximation of
Eq.~(\ref{equation_31}) has a limited range of validity. For
sigmoidal nonlinearities, as the firing rate increases, the
operation point $r_0^{\rm nl}$ shifts upwards, so the slope
$f'(q_0)$ should not be taken as fixed. The value of the
derivative diminishes progressively, as the flat part of the
nonlinearity is approached. In Eq.~(\ref{equation_34}), a
diminished $f'(q_0)$ is equivalent to a smaller feedback coupling
constant $g$, thereby precluding divergences. Hence, although the
firing probability can still contain a strong oscillatory
component in the nonlinear case, the increased firing at the peaks
of the oscillations acts as a self-regulatory mechanism, that
forestalls unstable dynamics.

In the present section, we have introduced a nonlinearity in the
relationship between $r(t)$ and $s_1(t)$. The feedback process,
however, always remained linear. One could then wonder what is the
effect of maintaining a linear relation between $r(t)$ and
$s_1(t)$, but introducing a nonlinearity in the feedback process.
From Eqs.~(\ref{equation_29}) and (\ref{equation_30}), we see that
in terms of $q(t)$, a nonlinear relation between $r(t)$ and
$s_1(t)$ traduces into a nonlinear system of equations, where the
nonlinearity appears in the feedback equation. Conversely, a
nonlinearity in the feedback process that modifies the second term
of the right-hand side of Eq.~(\ref{equation_8}) can be treated,
from the mathematical point of view, in the same way as the the
nonlinear system explored in this section. More general nonlinear
feedback processes, for example, modifying the first term of the
right-hand side of Eq.~(\ref{equation_8}), or even mixing the two
terms together, require additional techniques.

\subsection{Higher-dimensional receptive fields}
\indent The results obtained so far remain unchanged if the
stimulus (and therefore also the linear filter) depends on several
dimensions, and not just time, as long as adaptation is
homogeneous in those dimensions. In this case, we write the
external stimulus as $s({\bf y}, t)$, where ${\bf y} \in {\cal D}$
is a vector in a one- or a multi-dimensional space ${\cal D}$
representing the relevant features of the sensory modality under
study (direction of the incoming light or chromatic composition in
the case of vision, frequency content in the case of audition,
etc). The external stimulus is filtered by some sort of
transduction process and perhaps also by one or more neurons that
lie between the sensory receptors and the neuron under study. We
represent such upstream filtering processes by means of a
high-dimensional filter $h_u({\bf y},\tau)$, that transforms the
high-dimensional stimulus $s({\bf y}, t)$ in a purely temporal
ionic input current $s_1(t)$,

\begin{equation} \label{equation_38}
  s_1(t) = \int_{\cal D} \int_{-\infty}^{+\infty} ~h_u({\bf y},\tau)
  ~s({\bf y}, t - \tau) \ {\rm d}{\bf y} \ {\rm d}\tau.
\end{equation}

\noindent This expression can be inserted into
Eq.~(\ref{equation_2}) to construct the new full input current
$I(t)$ needed in Eq.~(\ref{equation_1}). Notice that in the
present framework, the high-dimensional stimulus is first filtered
in the ${\bf y}$-dimensional and temporal domains by $h_u({\bf y},
\tau)$ (Eq.~(\ref{equation_38})), and afterwards by the temporal
filter $h(\tau)$ (Eq.~(\ref{equation_1})). Using a Fourier
analysis completely analogous to the one developed before, we
arrive at

\begin{eqnarray} \label{equation_39}
  \hat{r}(\omega) &=& \frac{\sqrt{2\pi}~(1+i~\omega~\tau_{\rm d})}
  {1+i~\omega~\tau_{\rm d} + \sqrt{2\pi}~g~\tau_{\rm d}~
  \hat{h}(\omega)}~\Bigg[ r_0~\delta(\omega) \nonumber\\
  && \hspace{1cm} + ~\sqrt{2\pi}~\hat{h}(\omega)~\int_{\cal D}
  \hat{h_u}({\bf y},\omega)~\hat{s}({\bf y}, \omega)
  ~{\rm d}{\bf y} \Bigg].
\end{eqnarray}

\begin{sloppypar}
\noindent This expression reduces to Eq.~(\ref{equation_9}) when
$h_u({\bf y}, \tau) = \delta({\bf y} - {\bf y}_0) \ \delta(\tau)$.
\end{sloppypar}

\indent By stimulating the cell with signals that are localized in
the additional dimensions, $s({\bf y}, t) = \delta({\bf y} - {\bf
y}_0)~f(t)$, the whole of the previous theory becomes valid for
each chosen ${\bf y}_0$. In particular, feedback still transforms
the intrinsic receptive field in an effective receptive field by
multiplying the Fourier transform of the former by a factor that
depends on the frequency, but does not depend on the stimulation
point ${\bf y}_0$. At least, such is the effect of feedback if we
may assume that the signal $x(t)$ only depends on the output of
the cell and is not modulated, for example, by spatial input
components.

\section{Discussion and conclusions}

\indent  In this paper, we analyzed how spike-evoked negative
feedback modifies the effective receptive field of a cell. The
approach was based on an ideal concept, here named {\sl perfect
feedback}, where the signal $x(t)$ is not a function of the actual
spikes generated by the neuron, but rather of the {\sl
probability} that spikes be generated. This assumption is
ultimately unrealistic, but becomes a good approximation of the
real system when (a) the stimulus varies slowly compared to the
inter-spike interval of the neuron under study, or (b) feedback is
mediated by a large number of similar neurons in the network. In
both cases, the firing probability $r(t)$ is sampled exhaustively,
and therefore, the distribution of sampled signals follows the
probability $r(t)$ closely during the time scale $\tau_{\rm d}$
governing the feedback process. This idealized scenario allowed us
to develop an analytical approach, and to derive the mathematical
connection between the intrinsic filter $h(\tau)$ and the
effective filter $h^{\rm fb}(\tau)$. In particular,
Eq.~(\ref{equation_13}) provides the link by which intrinsic or
feedback parameters shape the receptive field.

\begin{sloppypar}
\indent Previous studies have reported some reshaping of receptive
fields as the input changes. For example, in the case of visual
stimuli, the spatial and the temporal context alter the input/output
transformation, as the system adapts to the local statistics
\cite{Schwartz_etal2007, Wark_etal2007, Wark_etal2009}, producing
changes all the way up to the perceptual level
\cite{Schwartz_etal2007, Kohn2007, Lochmann_etal2012}. In single
cells, both the total luminance and the amount of contrast shape
linear filters with increased band-pass characteristics
\cite{Enroth-CugellShapley1973, KaplanBenardete2001}. The same
effects are observed in single auditory cells, as the mean and the
variance of synthetic sounds are manipulated \cite{NagelDoupe2006}.
According to our results, these phenomena could be explained by
modulating the amount of feedback as the signal varies. The
modulation could be mediated by synaptic scaling, in the case of
network-based feedback processes, or by ionic mechanisms, in the
case of single cell adaptation, eg. by Ca$^{2+}$ concentration.
Moreover, the reshaping of filters can also be modulated by the
spatial structure of the stimulus. For example, the receptive fields
processing stimuli with natural spatio-temporal statistics are
different from those obtained from simple ensembles. When the
higher-order input statistics are taken into account in the
estimation of receptive fields \cite{Theunissen_etal2001}, the
temporal profile of the filters processing natural stimuli turn out
to be biphasic, whereas monophasic filters are obtained in responses
to gratings \cite{David_etal2004}. In terms of our analysis, this
change could be explained if the amount of negative feedback
depended on the statistical properties of the input signal. However,
given the different long-range behavior of spatial correlations for
different ensembles, this would require to extend the model in order
to include some spatial dependence of the feedback signal, for
example, associated to the spiking activity of neurons that respond
to stimuli presented in a shifted position.

\indent In the purely linear case, positive feedback always
produces unstable dynamics. If an ON cell is subject to positive
feedback, any positive stimulus fluctuation, no matter how small,
feeds a reverberating loop where activity eventually diverges. Any
negative stimulus fluctuation, in turn, eventually extinguishes
firing altogether. An OFF cell is also unstable, with the opposite
effect of positive and negative stimulus fluctuations. Such
unstable systems are all-or-none (divergence or extinction) and
have therefore not been studied here.

\indent An important finding of this paper is that negative
feedback, which is usually assumed to exert a regularizing effect,
can also produce unstable dynamics in the linear case. Negative
feedback always diminishes the responses to slow stimuli,
enhancing the band-pass characteristics of the filtering process.
However, in certain conditions, these characteristics can be
magnified dramatically, up to the point that the effective filter
$\hat{h}^{\rm fb}(\omega)$ be sharply peaked at a specific
frequency $\omega_0$. In the time domain, $h^{\rm fb}(\tau)$
exhibits pronounced oscillations at this particular frequency.
Feedback, hence, transformed an intrinsic receptive field that
acted as an integrator into an effective resonator
\cite{MatoSamengo2008}.

\indent Neurons often display resonant properties. A widely
accepted view states that oscillatory properties can either stem
from intrinsic cellular characteristics or from network
interactions \cite{Buzsaki2006, Prescott_etal2008, Wang2010}. The
important ingredient is that two types of mechanisms coexist
\cite{HutcheonYarom2000}: those attenuating high frequencies
(typically, leak currents), and those attenuating low frequencies.
Many processes can be invoked to attenuate low frequencies, both
at the single-cell (e. g., leak filtering), and the network level
(e. g., synaptic filtering). Previous studies of resonant behavior
have attempted to understand resonances in terms of such
properties, invoking particular subthreshold processes
\cite{Izhikevich2007, Gutfreund_etal1995, Hutcheon_etal1996a,
Hutcheon_etal1996b, Richardson_etal2003}, or specific network
interactions \cite{Wang2010, LamplYarom1997, Kondgen_etal2008,
BuzsakiWang2012}. In addition, to avoid falling into quiescence,
amplifying processes are sometimes also invoked. In agreement with
previous studies on properties of neurons with spike-frequency
adaptation \cite{BendaHerz2003, Gigante_etal2007, BendaHennig2008,
Benda_etal2010}, in this paper we have shown that negative
feedback suffices to attenuate low frequencies. We make no
assumptions about subthreshold properties, inasmuch as they
produce an intrinsic receptive field that contains a maximal
cutoff frequency. In addition, no ad-hoc mechanisms are required
to produce amplification, as long as the baseline firing rate
$r_0^{\rm fb}$ remains positive, for which a strong stimulus
baseline $s_0$, or a strong spontaneous rate $h_0$, suffice.
Moreover, the formalism proposed here is general enough as to be
equally applicable to network-mediated feedback currents, or to
intrinsic adaptation currents. Our idealized approach, hence,
provides a unified description of the mechanisms through which
spike-triggered negative feedback induces resonances.

\indent In order to test the validity of the idealized approach
provided by perfect feedback, we also ran numeric simulations
where a finite number $N$ of individual feedback signals $x_i(t)$
were triggered by actual spikes. Importantly, we concluded that
the effective filtering characteristics do not depend on $N$.
Perfect feedback, hence, can be safely used to study both
self-inhibition ($N = 1$) and network-induced regulation ($N >
1$). The analytic study of the fluctuations in the feedback signal
and in the response, however, shows that the amount of noise in
the output spectrum diminishes as $1/N$ (Eqs.~(\ref{equation_22})
and (\ref{equation_25})). Moreover, the shape of the noise
spectrum is given by a Cauchy distribution
\cite{Johnson_etal1994}.

\indent The above conclusions hold for linear Poisson neuron
models. Only for the linear case can the analytic approach be
developed. The concept of receptive field, however, can also be
extended to the case of linear-nonlinear Poisson models, or
generalized nonlinear models. Although, strictly speaking, the
presence of a nonlinearity does not allow us to employ linear
methods, in the last section of this paper we demonstrated that by
linearizing one of the two equations governing the system, an
approximate description of the nonlinear case is possible. We
showed that feedback still produces an effective receptive field
that is narrower than the intrinsic one, and that resonances may
also appear. Moreover, the nonlinear description is also useful to
understand how the divergences obtained in the purely linear case
saturate at a finite value in the nonlinear description.
\end{sloppypar}

\begin{acknowledgements}
This work has been funded by Consejo Nacional de Investigaciones
Cient\'ificas y T\'ecnicas, Agencia Nacional de Promoci\'on
Cient\'ifica y Tecnol\'ogica, Universidad Nacional de Cuyo and
Comisi\'on Nacional de Energ\'ia At\'omica, all from Rep\'ublica
Argentina.
\end{acknowledgements}

\appendix
\section{Appendix}\label{Apendice1}

\indent The convention used here to operate with the Fourier
transform is

\begin{equation}\label{A1}
\hat{f}(\omega) =
   \frac{1}{\sqrt{2\pi}}\int_{-\infty}^{+\infty}
   f(t)~{\rm e}^{-i\omega t}~{\rm d}t.
\end{equation}

\indent From this definition, the following properties follow:

\begin{sloppypar}
\begin{enumerate}
\item[-] The Fourier transform of a constant signal of magnitude
$r_0$ is $\hat{f}(\omega) = \sqrt{2 \pi} \  r_0 \ \delta(\omega)$.
\item[-] If a signal $f(t)$ is equal to the convolution of two
other signals $g(t)$ and $h(t)$, then $\hat{f}(\omega) = \sqrt{2
\pi} \ \hat{g}(\omega) \ \hat{h}(\omega)$. \item[-] If a signal
$f(t)$ is equal to another signal but delayed, $f(t) =
g(t-\Delta)$, then $\hat{f}(\omega) = {\rm e}^{-i \omega \Delta}
~\hat{g}(\omega)$. \item[-] If a signal $f(t)$ is the derivative
of another signal $f(t) = {\rm d}g/{\rm d}t$, then
$\hat{f}(\omega) = i \omega \hat{g}(\omega)$.
\end{enumerate}
\end{sloppypar}

\indent The Laplace transform, in turn, is defined as

\begin{equation}\label{A2}
\tilde{f}(s) = \int_{0}^{\infty}
   f(t)~{\rm e}^{-s t}~{\rm d}t,
\end{equation}

\noindent where $s = \sigma + i \omega$ is a complex number. In
the particular case where $s$ is evaluated at a purely imaginary
number ($\sigma = 0$), the Laplace transform is proportional to
the Fourier transform for temporally positive functions. The
Laplace transform, hence, can be seen as a generalization of the
Fourier transform to the whole complex plane. Related properties
hold:

\begin{enumerate}
\item[-] The Laplace transform of a constant signal of magnitude
$r_0$ is $\tilde{f}(s) = r_0 / s$. \item[-] If a signal $f(t)$ is
equal to the convolution of two other signals $g(t)$ and $h(t)$,
then $\tilde{f}(s) = \tilde{g}(s) \ \tilde{h}(s)$. \item[-] If a
signal $f(t)$ is the derivative of another signal $f(t) = {\rm
d}g/{\rm d}t$, then $\tilde{f}(s) = s \ \tilde{g}(s) - g(0)$.
\end{enumerate}

\indent Using these properties, it is easy to see that if the
signals $r(t)$ and $x(t)$ are governed by Eqs.~(\ref{equation_4})
and (\ref{equation_8}), the Laplace transform of $r(t)$, for an
initial zero feedback contribution $x(0) = 0$, is

\begin{equation}\label{A3}
\tilde{r}(s) = \frac{r_0 / s + \tilde{h}(s) \ \tilde{s}_{1}(s) }
{1 + g \ \tilde{h}(s) / (s + 1/\tau_{\rm d})}.
\end{equation}

\indent According to control theory \cite{Franklin_etal1994}, this
feedback system becomes unstable when at least one pole of
$\tilde{r}(s)$ has positive real part. It is important to search
for instabilities in the Laplace representation, since they may
not be evident in the Fourier space. In addition to the fixed pole
$s=0$ given by the constant signal term, the poles of
$\tilde{r}(s)$ are those complex points $s$ where the denominator
of Eq.~(\ref{A3}) vanish, that is,

\begin{equation} \label{A4}
 1 + g \frac{\tilde{h}(s)}{s + 1/\tau_{\rm d}} = 0.
\end{equation}

\indent This equation holds in the complex plane, so both the real
and the imaginary part of the equality must vanish. Feedback gives
rise to unstable behavior whenever at least one solution of
Eq.~(\ref{A4}) has positive real part. The onset of instability,
hence, appears when the pole (or pair of conjugate poles) with
largest real part crosses the imaginary axis, from left to right.
At the crossing, $\sigma = 0$, so the real and imaginary parts of
Eq.~(\ref{A4}) become

\begin{eqnarray}
  \label{A5}
  {\rm Im}[\hat{h}(\omega)] - \tau_{\rm d}~\omega~{\rm Re}
  [\hat{h}(\omega)] &=& 0, \\
  \label{A6}
  1 + \sqrt{2\pi}~g~\tau_{\rm d}~{\rm Re}
  [\hat{h}(\omega)] &=& 0,
\end{eqnarray}

\noindent where now, $\hat{h}(\omega)$ is the Fourier transform of
the intrinsic filter, obtained when evaluating the Laplace
transform at a purely imaginary point and dividing by
$\sqrt{2\pi}$. The frequencies satisfying Eq.~(\ref{A5}) are
independent of the feedback strength, $g$. Note that, in addition,
Eq.~(\ref{A6}) can only be fulfilled by those frequencies
$\omega_i$, from the set of solutions to Eq.~(\ref{A5}), that
result in ${\rm Re}[\hat{h}(\omega_i)] < 0$. Even among these, if
$g$ is small, no frequency $\omega$ satisfies Eq.~(\ref{A6}), and
the system is stable (the pole with largest real part is on the
stable semi-plane). For a critical value of feedback strength, the
condition imposed by Eq.~(\ref{A6}) can be finally reached and the
system becomes unstable. This means that for feedback strengths
beyond the critical value, there is at least one pole on the
unstable semi-plane and therefore the inverse Fourier transform
diverges.

\section{Appendix}\label{Apendice2}
\begin{sloppypar}
\indent In this section, we describe the temporal
processing of slow stimuli. We arrive at the same results as the
ones derived by \cite{UrdapilletaSamengo2009}, but here we base
the analysis on the spectral properties of filters. In the absence
of feedback,
\end{sloppypar}

\begin{equation}
   r(t) = h_0 + \int_{-\infty}^{+\infty} h(\tau)~s(t-\tau)~{\rm
   d}\tau, \nonumber
\end{equation}

\begin{equation}\label{B1}
   \hat{r}(\omega) = \sqrt{2\pi}\left[ h_0~\delta(\omega) +
   \hat{h}(\omega)~\hat{s}(\omega) \right].
\end{equation}

\indent In the present context, a slow stimulus is one that does
not contain high frequency components. In other words, the
input/output relation of the cell is only determined by the lowest
frequency components of the filter $\hat{h}(\omega)$, since the
higher frequencies are not explored. The stimulus, hence, has to
remain fairly constant throughout the time scale of the filter
(given by the non-zero portions in the temporal filters shown in
Fig.~\ref{fig1}). If only low frequencies matter, we may take the
limit $\omega \to 0$. In this context, we prove that\\

\begin{itemize}
\item ON cells behave as low-pass filters, see Fig.~\ref{fig1}A.
For $\omega \rightarrow 0$, the Fourier transform of these filters
is

\begin{eqnarray}
   \hat{h}(\omega) &=& \frac{1}{\sqrt{2\pi}} \int_{-\infty}^{+\infty}
   h(\tau)~{\rm e}^{-i \omega \tau}~{\rm d}\tau \nonumber\\
   &=& \frac{1}{\sqrt{2\pi}} \int_{-\infty}^{+\infty}
   h(\tau)~[\cos(\omega \tau)-i~\sin(\omega \tau)]~{\rm
   d}\tau\nonumber\\
   \label{B2}
   &\approx& \frac{1}{\sqrt{2\pi}} \int_{-\infty}^{+\infty} h(\tau)~[1 -
   i~\omega~\tau]~{\rm d}\tau = \frac{1}{\sqrt{2\pi}} \left( H
   + i~\omega~H_1 \right)
\end{eqnarray}

\noindent where $H = \int_{-\infty}^{+\infty} h(\tau)~{\rm d}\tau$
and $H_1 = - \int_{-\infty}^{+\infty} ~\tau ~h(\tau)~{\rm d}\tau$.
The small angle approximation can be used again,

\begin{eqnarray}
   \hat{h}(\omega) &\approx& \frac{H}{\sqrt{2\pi}} \left[
   1 + i~\omega~\frac{H_1}{H} \right] \nonumber\\
   &\approx& \frac{H}{\sqrt{2\pi}} \left[ \cos\left(\omega ~
   \frac{H_1}{H}\right) + i~\sin\left(\omega~
   \frac{H_1}{H}\right)\right] \nonumber\\
   \label{B3}
   &\approx& \frac{H}{\sqrt{2\pi}}~{\rm e}^{-i \omega
   \delta_{0}},
\end{eqnarray}

\noindent where we have defined the (positive) delay $\delta_{0} =
-H_1/H$. From this expression, it is easy to check that the gain
of the Fourier transform for ON cells at low frequencies is
constant and equal to $H/\sqrt{2\pi}$. In addition, the phase
decreases linearly with $\omega$, starting at $0$ with slope
$-\delta_0$ (in Fig.~\ref{fig1}A, a linear graph instead of the
semi-logarithmic one would clearly show this linear relationship).

\indent To relate $\hat{h}(\omega)$ to the neural response in the
temporal domain, we simply make use of the Fourier transform of a
delayed signal (see Appendix \ref{Apendice1}),

\begin{equation} \label{B4}
   r(t) \approx h_0 + \mathcal{F}^{-1} \Big\{ H~{\rm e}^{-i \omega
   \delta_{0}}~\hat{s}(\omega)\Big\} \approx h_0 + H ~s(t-\delta_{0}).
\end{equation}

\indent According to Eq.~(\ref{B4}), the amplitude of the response
is exclusively determined by $H$; simultaneously, neural
processing introduces a fixed delay $\delta_{0}$ between the
response and the stimulus.\\

\item ON biphasic cells behave as band-pass filters, see
Fig.~\ref{fig1}C. In the limit $\omega \rightarrow 0$,
$\hat{h}(\omega)$ is still given by Eq.~(\ref{B1}). However,
symmetric biphasic filters satisfy $H = 0$, so to obtain a
meaningful description we have to perform the expansion around
$\omega = 0$ up to the second order,

\begin{eqnarray}
   \hat{h}(\omega) &\approx& \frac{1}{\sqrt{2\pi}} \int_{-\infty}^{+\infty}
   h(\tau)~\left[1 - i~\omega~\tau - \frac{1}{2} \omega^{2} \tau^{2}\right]
   ~{\rm d}\tau \nonumber\\
   &\approx& \frac{1}{\sqrt{2\pi}} \left( \cancel{{H}} + i~\omega~
   H_1 - \frac{1}{2} ~\omega^{2}~H_2 \right),\nonumber\\
   \label{B5}
   &\approx& \frac{i~H_1~\omega}{\sqrt{2\pi}}\left[
   1 + i~\omega~\frac{H_2}{2H_1}\right].
\end{eqnarray}

\noindent where $H_2 = \int_{-\infty}^{+\infty} ~\tau^2
~h(\tau)~{\rm d}\tau$. Proceeding as before and defining a
corresponding (positive) delay $\delta_{1} = -H_2/(2H_1)$, we
obtain

\begin{eqnarray}
   \hat{h}(\omega) &\approx& \frac{H_1~\omega}{\sqrt{2\pi}}~
   {\rm e}^{i \pi/2} ~\left[ \cos\left( \omega ~\frac{H_2}
   {2H_1}\right) + i ~\sin\left(\omega ~\frac{H_2}
   {2H_1}\right)\right]\nonumber\\
   \label{B6}
   &\approx& \frac{H_1~\omega}{\sqrt{2\pi}}~{\rm e}^{i \pi/2}
   ~{\rm e}^{-i \omega \delta_{1}} = \frac{H_1~\omega}{\sqrt{2\pi}}
   ~{\rm e}^{i(\pi/2-\omega \delta_{1})}.
\end{eqnarray}

\indent Hence, in the limit $\omega \rightarrow 0$, the gain of
symmetric biphasic filters depends linearly on the angular
frequency, with a positive slope $H_1/\sqrt{2\pi}$. In addition,
the filter phase is a linearly decreasing function of $\omega$,
with intercept at the origin $\pi/2$ and slope $-\delta_{1}$. Both
characteristics can be observed in Fig.~\ref{fig1}C.

\indent To relate $\hat{h}(\omega)$ to the neural response in the
temporal domain, it is useful to recover the imaginary unit as a
factor and rewrite Eq.~(\ref{B6}) as

\begin{eqnarray}\label{B7}
   \hat{h}(\omega) &\approx& \frac{H_1}{\sqrt{2\pi}}~
   {\rm e}^{-i \omega \delta_{1}}~(i~\omega).
\end{eqnarray}

\indent In this case, given that the Fourier transform of a
convolution becomes a product in Fourier space, the term
$(i~\omega)$ can be effectively associated to the stimulus,
$\hat{s}(\omega)$. In turn, by applying the Fourier transform of a
derivative, and proceeding as before, it is easy to check that
$r(t)$ is proportional to the delayed stimulus derivative, with a
factor of proportionality given by $H_1$ and a fixed delay
$\delta_{1}$. Explicitly,

\begin{equation}\label{B8}
   r(t) \approx h_0 + \mathcal{F}^{-1}\Big\{ H_{1}~{\rm e}^{-i \omega
   \delta_{1}} ~\left[i~\omega~\hat{s}(\omega) \right]\Big\}
   \approx h_0 + H_1 \left[ \frac{{\rm d}s}{{\rm d}t} \right]_{(t-\delta_{1})}.
\end{equation}

\item Corresponding OFF cells behave analogously to monophasic and
biphasic ON cells (see Figs.~\ref{fig1}B and \ref{fig1}D).
Specifically, the Fourier transforms of these filters are exactly
given by Eqs.~(\ref{B2}) and (\ref{B5}), in the limit $\omega
\rightarrow 0$. The only difference with the previous ON cells is
that the proportionality factors $H$ and $H_1$ for monophasic and
biphasic cells, respectively, are now negative. The presence of a
factor $(-1)$ affecting the whole expression can be incorporated
into a multiplicative term, $\exp{(\pm i \pi)}$, which shifts the
phase of the Fourier transform in $\pm \pi$. This shift can be
observed in Fig.~\ref{fig1} by comparing phases of corresponding
filters (monophasic ON/OFF filters and biphasic ON/OFF filters).
Additionally, the factor $(-1)$ does not affect the magnitude of
the Fourier transform of ON or OFF cells. In the temporal domain,
this factor simply means that the relationships obtained for ON
cells are also valid, but associated to the negative of the
stimulus or its derivative. These results agree our previous
analysis \cite{UrdapilletaSamengo2009}.
\end{itemize}

\end{document}